\documentclass[%
 aip,
 sd,%
 amsmath,amssymb,
 reprint,%
]{revtex4-1}
\def\PT{$\cal{PT}$}
\def\[{\begin{equation}}
\def\]{\end{equation}}
\usepackage{graphicx}
\usepackage{dsfont}
\usepackage{dcolumn}
\usepackage{bm}

\begin{document}

\preprint{AIP/123-QED}

\title[Rogue Waves in the Reverse Time Integrable Nonlocal Nonlinear Equations]{General Rogue waves and their dynamics in several \\
reverse time integrable nonlocal nonlinear equations}
\author{Bo Yang}
\affiliation{
Shanghai Key Laboratory of Trustworthy Computing, East China Normal University, Shanghai, 200062, People's Republic of China
}%
\affiliation{%
MOE International Joint Lab of Trustworthy Software, East China Normal University, Shanghai, 200062, People's Republic of China
}%

\author{Yong Chen}%
 \email{(Corresponding author.)\ ychen@sei.ecnu.edu.cn.}
\affiliation{
Shanghai Key Laboratory of Trustworthy Computing, East China Normal University, Shanghai, 200062, People's Republic of China
}%
\affiliation{%
MOE International Joint Lab of Trustworthy Software, East China Normal University, Shanghai, 200062, People's Republic of China
}%
\affiliation{Department of Physics, Zhejiang Normal University, Jinhua, 321004, China}

\begin{abstract}
A study of general rogue waves in some integrable reverse time  nonlocal nonlinear equations is presented. Specifically, the reverse time  nonlocal nonlinear Schr\"odinger (NLS)  and   nonlocal Davey-Stewartson (DS) equations  are investigated,  which  are nonlocal reductions from the  AKNS hierarchy. By using  Darboux transformation (DT) method, several types of rogue waves are constructed.  Especially, a unified binary DT is found for this nonlocal DS system, thus the solution formulas for nonlocal DSI and DSII equation can be written  in an uniform expression.  Dynamics of these rogue waves is separately explored. It is shown that the  (1+1)-dimensional rogue waves in nonlocal NLS equation can be bounded for both $x$  and $t$, or develop collapsing singularities.  It is also shown that the (1+2)-dimensional line rogue waves in the nonlocal DS equations  can be bounded for all space and time, or have finite-time blowing-ups. All these types depend on the  values of  free parameters introduced in the solution. In addition, the dynamics patterns in the multi- and higher-order rogue waves exhibits more richer structures, most of which have no counterparts in the corresponding local nonlinear equations.
\end{abstract}

\maketitle

\section{Introduction}
The integrable nonlinear evolution equations are exactly solvable models which play an important role in the field of nonlinear science, especially in the study of nonlinear physical systems, including nonlinear optics, Bose-Einstein condensates, plasma physics  and ocean water waves. Most of these integrable equations are local equations,  that is,  the solution¡¯s  evolution only depends on the local solution value. In recent years, numbers of new  integrable nonlocal equations were proposed and studied\cite{AblowitzMussPRL2013,AblowitzMussPRE2014, AblowitzMussSAPM, Khara2015, YanZY2015,AblowitzMussNonli2016,ZhuRTSasa2017,BoTransformations,Fokas2016,Lou,Lou2,ZhoudNLS,ZhouDS,HePPTDS,Zhu1,Zhu2,Gerdjikov2017,Ablowitz_arxiv,JBY2017,BYnonlocalDS}. The first such nonlocal  equation was the $\mathcal{PT}$-symmetric nonlocal nonlinear Schr\"odinger (NLS) equation\cite{AblowitzMussPRL2013}:
\[ \label{e:PTNLS}
\textrm{i} q_t(x,t)=q_{xx}(x,t)+2\sigma q^2(x,t)q^*(-x,t).
\]
Here, $\sigma=\pm 1$ is the sign of nonlinearity (with the plus sign being the focusing case and minus sign the defocusing case),  and the asterisk * represents complex conjugation.   It is noted that  $\mathcal{PT}$-symmetric systems have attracted a lot of attention in optics and other physical fields in recent years\cite{Kivsharreview,Yangreview,PTNLSmagnetics,KKTLPRL2015, BKBPRA20152015}.

Following this nonlocal $\mathcal{PT}$-symmetric NLS equation, some new reverse space-time and reverse time type nonlocal nonlinear integrable equations were also introduced and quickly reported\cite{AblowitzMussNonli2016,ZhuRTSasa2017,BoTransformations}. They are integrable infinite dimensional Hamiltonian dynamical systems,  which arise from remarkably simple symmetry reductions of general ZS-AKNS scattering problems where the nonlocality appears in both space and time or time alone.  These examples including the  reverse space-time and/or the reverse time only nonlocal NLS\cite{AblowitzMussNonli2016}, the (complex)  modified Korteweg-deVries, sine-Gordon,  three-wave interactions, derivative NLS (dNLS),  Davey-Stewartson, discrete NLS type\cite{AblowitzMussNonli2016}, Sasa-Satsuma\cite{ZhuRTSasa2017}(S-S) and nonlocal complex short pulse\cite{BoTransformations}(CSP) equations , and many others\cite{BoTransformations}.

These nonlocal equations, because of their novel space and/or time coupling, are distinctly different from local equations. Indeed, solution properties in some of these nonlocal equations have been analyzed by the inverse scattering transform method, Darboux transformation or the bilinear method.  On the one hand, those new systems could reproduce  solution patterns  which  already have been discovered in their local counterparts.  On the other hand,  some interesting behaviors such as blowing-up solutions with finite-time singularities\cite{BoTransformations, BYnonlocalDS}  and the  existence of even more richer structures have also been revealed\cite{ZhuRTSasa2017, HePPTDS,Zhu1,BYnonlocalDS}. Although these  nonlocal equations are mathematically interesting. In the view of further potential applications,  it links to an unconventional system of magnetics\cite{PTNLSmagnetics},  and relates to the concept of $\mathcal{PT}$-symmetry,  which is a  hot research area in contemporary physics\cite{Yangreview}.

Rogue waves  have  attracted a lot of attention in recent years due to their dramatic and often damaging effects, such as in the ocean and optical fibers \cite{KPS2009,SRKJ2007}.   The first analytical expression of a rogue wave for the NLS equation was derived by Peregrine in 1983 \cite{PDH1983}. Later, the analytical rogue-wave solutions have been derived and  interesting dynamical patterns been revealed for a large number of integrable systems\cite{Akhmediev2009,AAS2009,ACA2010,DGKM2010,DPMB2011,GLML2012,OhtaJY2012,XuHW2011,BCDL2013,
OhtaJKY2012,OhtaJKY2013,AANJM2010,OhtaJKY2014,ASAN2010,TaoHe2012,BDCW2012,PSLM2013,Grimshaw_rogue,MuQin2016,LLMSasa2016,LLMFZ2016,XinABsys2015, JCChen2017, XenYongKdV2017}.

As an unexplored  and interesting subject, rogue waves in the  nonlocal integrable systems have received much attention.  For local integrable equations, the evolution for most rogue-wave  solutions depends only on the local solution value with its local space and time derivatives. However, for the nonlocal equations,  the sates of rogue-wave solutions at distinct locations $x$  and $-x$  are directly related \cite{HePPTDS,BYnonlocalDS,JBY2017}.  Hence, these facts are basically important and further  motivate us ask an interesting open question: If one only consider the connections between the  sates of rogue-wave solutions at reverse time points $t$  and $-t$,  whether there are rogue waves existing in some reverse time  nonlocal nonlinear equations?

In this article,  to give an answer to this question,  we study rogue waves in several reverse time integrable nonlocal nonlinear equations.  As typically concrete  examples,  we focus on the reverse time nonlocal NLS equation:
\[ \label{e:RTNLS}
\textrm{i}q_t(x,t)=q_{xx}(x,t)+2\sigma q^2(x,t) q(x,-t),
\]
and the reverse time nonlocal DS equations:
\begin{eqnarray}
  && iq_{t}+\frac{1}{2}\gamma^{2} q_{xx}+\frac{1}{2} q_{yy}+ (q r - \phi)q=0, \label{PTDS} \\
  && \phi_{xx}-\gamma^{2} \phi_{yy} - 2\left(q r\right)_{xx}=0,\label{PTDSS}
\end{eqnarray}
where $r(x,y,t)=\sigma q(x,y,-t)$,   $q$, $r$ and $\phi$ are functions of $x,y,t$, $\gamma^2=\pm1$ is the equation-type parameter (with $\gamma^2=1$ being the DS-I and $\gamma^2=-1$ being DS-II). With certain reductions, Eq.(\ref{e:RTNLS})  and Eqs.(\ref{PTDS})-(\ref{PTDSS}) can be derived from  the member of the (1+1)- and (1+2)-dimensional AKNS hierarchy, respectively.

By using Darboux transformation method,  we derive general rogue waves for these three nonlocal equations,  solution formulas  are given under certain reductions of wave functions and adjoint wave functions.   More interestingly and coincidentally, we find a unified binary DT for this nonlocal DS system, so that rogue-wave solutions in nonlocal DSI and DSII equation can be expressed in a unified form, which is quite different from  the constrictions of  DT  in the local DS~\cite{MatveevDT1991} and partially $\mathcal{PT}$-symmetric DS equations\cite{BYnonlocalDS}.   In addition, dynamics of these rogue waves is further analyzed.  For these  reverse time nonlocal equations,  it is shown that general rogue waves can be bounded for all space and time. More importantly, they can also develop collapsing singularities (for nonlocal NLS)  or  finite time blowing-ups (for nonlocal DS), which have no counterparts for the local equations. In addition, under certain parameter conditions, the dynamics of multi-rogue waves and higher-order rogue waves can exhibit more patterns. Most of them haven't been found before in the integrable nonlocal  nonlinear equations.
\section{Rogue waves in the  reverse time nonlocal nonlinear Schr\"{o}dinger equation}
In this section, we consider rogue waves in the focusing reverse time nonlocal NLS equation (\ref{e:RTNLS}) (with $\sigma=1$),  which approach the unit constant background when $x, t\to \pm \infty$. In the local NLS equation, general rogue waves in the form of rational solutions have been reported in \cite{AAS2009,ACA2010,DGKM2010,DPMB2011,GLML2012,OhtaJY2012}, which are bounded solutions for both $t$  and $x$.   For the nonlocal NLS equation (\ref{e:PTNLS}), we will show that the reverse time nonlocal NLS equation admits a wider variety of rogue waves. In addition to the  non-collapsing patterns, there are also other types of rogue waves which can develop collapsing singularities.

It has been pointed that Eq.(\ref{e:RTNLS}) is an integrable Hamilton evolution equation that admits an infinite number of conservation laws~\cite{AblowitzMussNonli2016}.  The first four conserved quantities are verified and given by
\begin{eqnarray*}
&&   I_1=\int_{-\infty}^\infty q(x,t)q(x, -t)dx, \\
&&   I_2=\int_{-\infty}^\infty q(x, -t) q_x(x,t)dx, \\
&&   I_3=\int_{-\infty}^\infty \left[q(x, -t)q_{xx}(x,t)+ q^2(x,t)q^{2}(x, -t)\right] dx, \\
&&   I_4=\int_{-\infty}^\infty q(x, -t)\left\{q_{xxx}(x,t)+[q^2(x,t)q(x, -t)]_x  \right.   \\
&&   \hspace{3.0cm} \left. +2 q(x,-t)q(x,t) q_x(x,t)\right\}dx.
\end{eqnarray*}

For this equation,  the  evolution at time $t$ depends on not only the local solution at $t$, but also the nonlocal solution at the reverse time point $-t$. That is, solution states at reverse time points $t$ and $-t$ are directly related.  Especially, N-solitons are derived recently \cite{JYRTNLS2017} via the Riemann-Hilbert solutions of AKNS hierarchy under the reverse-time  nonlocal reduction.  In this part, we focus on the general rogue  waves.

We begin with the following ZS-AKNS scattering problem\cite{Ablowitz1981,Zakharov1984}:
\begin{eqnarray}
 &&\Phi_{x}=U(q, r, \lambda) \Phi,  \label{Laxpairxp}\\
 &&\Phi_{t}=V(q, r, \lambda) \Phi,  \label{Laxpairtp}
\end{eqnarray}
where $\Phi$ is a column-vector,
\begin{eqnarray}
&& U(q,r,\lambda)=  -\textmd{i} \lambda \sigma_{3} + Q ,  \label{Uform}\\
 && V(q,r,\lambda)= 2\textmd{i} \lambda^2 \sigma_{3} - 2\lambda Q  - \textmd{i} \sigma_{3} \left( Q_{x}-Q^2 \right), \label{Vform} \\
 \vspace{0.2cm}
&&  \sigma_{3}=\textmd{diag}(1,-1),\
Q(x,t)=\left(
                                         \begin{array}{cc}
                                           0 & q(x,t) \\
                                           r(x,t) & 0 \\
                                         \end{array}
                                       \right),
\  x,t \in \mathbb{R}. \nonumber
\end{eqnarray}
The compatibility condition of these equations give rise to the zero-curvature equation
\[ \label{zero_curvature}
U_t-V_x+[U, V]=0,
\]
which yields the following coupled system for potential functions $(q, r)$ in the matrix $Q(x,t)$:
\begin{eqnarray}
&& \textmd{i}q_t=q_{xx}-2q^2r, \label{qequation} \\
&& \textmd{i}r_t=-r_{xx}+2r^2q. \label{requation}
\end{eqnarray}
The focusing reverse-time nonlocal NLS equation (\ref{e:RTNLS}) is obtained from the above coupled system (\ref{qequation})-(\ref{requation}) under the symmetry reduction:
\[ \label{qS}
r(x,t)= -q(x,-t).
\]
As it is stressed by Ablowitz and Musslimani\cite{AblowitzMussSAPM}:`` This reduction is new, remarkably simple, which has not been noticed in the literature  and leads to a nonlocal in time NLS hierarchy."
Under reduction (\ref{qS}), the potential matrix $Q$ satisfies the following symmetry condition:
\[ \label{QS}
Q^T(x,t)=-Q(x,-t).
\]

To construct the Darboux transformation, we need to introduce the adjoint-spectral  problem for (\ref{Laxpairxp})-(\ref{Laxpairtp}):
\begin{eqnarray}
&& -\Psi_{x}= \Psi U(q,r,\zeta),\label{AdLaxpairxp} \\
&&  -\Psi_{t}= \Psi V(q,r,\zeta),\label{AdLaxpairtp}
\end{eqnarray}
here $\Psi$ is a row-vector.

\subsection{$N$-fold Darboux transformation and reverse-time reduction}
For the ZS-AKNS scattering problem, there is a general Darboux transformation proposed  in \cite{ManasDTNLS1996,Cieslinski2009}, which can be represented as:
\begin{eqnarray}\label{eDT}
T=I+\frac{\zeta_{1}-\lambda_{1}}{\lambda-\zeta_{1}}P_{1},\ P_{1}=\frac{\Phi_{1}\Psi_{1}}{\Psi_{1}\Phi_{1}},
\end{eqnarray}
where $\Phi_{1}=(\phi_{1}, \phi_{2})^T$ is a solution with spectral parameter $\lambda=\lambda_{1}$, and $\Psi_{1}=(\psi_{1}, \psi_{2})^T$ is a solution of conjugation system with spectral parameter $\zeta=\zeta_{1}$. This DT can convert  (\ref{Laxpairxp})-(\ref{Laxpairtp}) into a new  system
\begin{eqnarray*}\label{New-Lax-pair}
&&\left(\Phi_{[1]}\right)_{x}=U(q_{[1]},r_{[1]},\lambda) \Phi_{[1]}, \\
&& \left(\Phi_{[1]}\right)_{t}=V(q_{[1]},r_{[1]},\lambda) \Phi_{[1]},
\end{eqnarray*}
with the transformation between  potential matrix:
\[ \label{DTP}
Q_{[1]}=Q+\textrm{i}(\zeta_{1}-\lambda_{1})\left[  \sigma_{3}, P_{1}  \right].
\]

Next, we consider the reduction of DT for system (\ref{Laxpairxp})-(\ref{Laxpairtp}).  Due to the potential  symmetry (\ref{QS}), it is  shown by direct calculation that matrix $U$  satisfy the symmetry:
\begin{eqnarray}
 U^{T}(x,t,\lambda) = -U(x,-t,-\lambda).  \label{UVS1}
\end{eqnarray}
For the given form (\ref{Uform})  in matrix $U$, the matrix $V$ which satisfies the zero-curvature equation (\ref{zero_curvature}) with the specific form of $\lambda$-dependence as in Eq. (\ref{Vform}) is unique. Therefore, by utilizing symmetry (\ref{UVS1}) and the zero-curvature equation (\ref{zero_curvature}), we can derive the corresponding symmetry of the matrix $V$:
\begin{eqnarray}
 V^{T}(x,t,\lambda) = V(x,-t,-\lambda).
\end{eqnarray}

Using these $U$ and $V$ symmetries, we can derive the symmetries of wave functions $\Phi$ and adjoint wave functions $\Psi$, and hence the symmetry of the Darboux transformation for the reverse-time nonlocal NLS equation (\ref{e:PTNLS}). Applying these symmetries to the spectral problems (\ref{Laxpairxp})-(\ref{Laxpairtp}), we get
\begin{eqnarray}\label{WFreduction}
&&  - \Phi_{x}^{T}(x,-t) = \Phi^{T}(x,-t) U(x,t,-\lambda),\\
&&  - \Phi_{t}^{T}(x,-t)=  \Phi^{T}(x,-t) V(x,t,-\lambda).
\end{eqnarray}
Thus, if $\Phi(x,t)$ is a wave function of the linear system (\ref{Laxpairxp})-(\ref{Laxpairtp}) at $\lambda$, then $ \Phi^{T}(x,-t)$ is an adjoint wave function of the  adjoint system at $\zeta=-\lambda$. In this case, if $\Psi_1(x,t,\lambda)$  and $ \Phi^{T}(x,-t,-\lambda)$ are linearly dependent on each other, then matrix (\ref{eDT}) would preserve the potential reduction (\ref{qS}) and thus be a Darboux transformation for the reverse-time nonlocal NLS equation (\ref{e:PTNLS}). Specifically, we have the following results.

\textbf{Proposition 3.1}.   For any spectral $\lambda_{1}, \zeta_{1} \in  \mathbb{C}$,  if
\[ \label{PhiPsicond}
\zeta_{1}=-\lambda_{1},\ \Psi_1(x,t,\zeta_{1})=\alpha \hspace{0.05cm} \Phi^{T}_1(x,-t,\lambda_{1}),
\]
where $\alpha$  is a  complex constant. Then the Darboux matrix (\ref{eDT}) accords with a Darboux transformation for
the focusing reverse-time nonlocal NLS equation (\ref{e:PTNLS}).

This proposition can be readily proved by checking that the new potential matrix $Q_{[1]}$ from Eq. (\ref{DTP}) satisfies the symmetry (\ref{qS}) under conditions (\ref{PhiPsicond}).
Via a direct calculation, this property can be easily verified.

The N-fold Darboux transformation is a  $N$ times iteration of the elementary DT with corresponding reductions. For the local  NLS equation, a N-fold Darboux matrix  has been given in \cite{BGLM2015}. Here, with  the reduction given above, we have the following N-fold Darboux matrix.

\textbf{Proposition 3.2}.  \emph{The N-fold  Darboux transformation matrix for the focusing \PT-symmetric NLS equation  can be represented as:}
\begin{eqnarray}\label{DT}
  T_{N}=I- Y M^{-1}D^{-1} X,
\end{eqnarray}
\emph{where},
\begin{eqnarray*}\nonumber
&&Y=\left[\  | y_{1} \rangle,\ | y_{2}  \rangle,\ldots , |y_{N} \rangle \ \right], \  |y_{k}\rangle=\Phi_{k}(\lambda_{k}),\\
&&X=\left[\  \langle x_{1}|,\ \langle x_{2}|,\ldots , \langle x_{N} | \ \right],\    \langle x_{k}|=\Psi_{k}(\zeta_{k}),\\
&&M=\left(m^{(N)}_{i,j}\right)_{1\leq i,j \leq N}, \ m^{(N)}_{i,j}=\frac{\langle x_{i}| y_{j}\rangle }{\lambda_{j}-\zeta_{i}},\\
&&D=\textrm{diag}\left( \lambda-\zeta_{1},  \lambda-\zeta_{2},\ldots,  \lambda-\zeta_{N}\right),
\end{eqnarray*}
\emph{where $|y_{i}\rangle$ solves the spectral equation (\ref{Laxpairxp})-(\ref{Laxpairtp}) at $\lambda=\lambda_{i}$, and $\langle x_{j}|$ solves the adjoint spectral equation (\ref{AdLaxpairxp})-(\ref{AdLaxpairtp}) at $\zeta=\zeta_{j}$.}
\emph{Here, for any $k \in \mathds{N}^{+}$, we should have}\ $\zeta_{k}=-\lambda_{k}$ \emph{with}$\langle x_{k}(x,t) |=|y_{k}(x,-t)\rangle^{T}$.

\emph{Moreover, the B\"{a}cklund  transformation between potential functions is:}
\begin{eqnarray}\label{gDTpotential}
&& u^{[N]}=
u+
 2\textmd{i}\frac{\left|
\begin{array}{cc}
 M &  X_{2} \\
 Y_{1} & 0 \\
\end{array}
\right|}
{\left| M \right|},
\end{eqnarray}
\emph{where $Y_1$ represents the first row of matrix $Y$, and $X_{2}$ represents the second column of matrix $X$.}

Expression (\ref{gDTpotential}) can be found\cite{Yang2010}, which appears as Riemann-Hilbert solution. The proof of this theorem has already been given.\cite{Yang2010,BGLM2015}

\subsection{\label{sec:level2}Derivation of general rogue-wave solutions}
In this section, we derive a general formula for rogue waves. First of all, we need the general eigenfunctions solved from the linear system
(\ref{Laxpairxp})-(\ref{Laxpairtp}) and its adjoint system (\ref{AdLaxpairxp})-(\ref{AdLaxpairtp}). Choosing a plane wave solution $u_{[0]}=e^{-2 \textmd{i} t}$ to be the seed solution, we can derive a general wave function for the linear system
(\ref{Laxpairxp})-(\ref{Laxpairtp}) as
\[ \label{Spectraleigen0}
\Phi(x,t)=\mathcal{D}\phi(x,t),
\]
where $\mathcal{D}=\mbox{diag}\left(e^{-\textmd{i}t}, e^{\textmd{i}t}\right)$,
\begin{eqnarray}\label{Spectraleigen}
&&\phi(x,t)=\left(
  \begin{array}{c}
    c_{1}e^A+c_{2}e^{-A}  \\
    c_{4}e^A+c_{3}e^{-A} \\
  \end{array}
\right), \\
&& A=\sqrt{-\lambda^2-1}(x-2 \lambda t+\theta), \nonumber \\
&& c_{3}=c_{2} \left(\textrm{i} \lambda-\sqrt{-\lambda^2-1}\right), \  c_{4}= c_{1}\left(\textrm{i} \lambda+\sqrt{-\lambda^2-1}\right),  \nonumber
\end{eqnarray}
and $c_{1}$, $c_{2}$, $\theta$ are arbitrary complex constants. Imposing the conditions of
$\lambda$ being purely imaginary, i.e., $\lambda=ih$, $|\lambda|>1$, $\theta$ is complex.
Furthermore,  to obtain the adjoint wave function, one can use  symmetry condition (\ref{PhiPsicond}). Through a simple Gauge transformation on the complex constant $c_1$, $c_{2}$, we can normalize $\alpha=1$.   Hence, the adjoint wave function to Eqs.(\ref{AdLaxpairxp})-(\ref{AdLaxpairtp}) at $\zeta=-\lambda$ satisfying (\ref{PhiPsicond})  can be given as:
\[
\Psi(x,t,\zeta)=\psi(x,t,\zeta)\mathcal{D}^{*},
\]
where
\begin{eqnarray}\label{Adeigen}
&&\psi(x,t,\zeta)=\phi^T(x,-t,\lambda).
\end{eqnarray}
Moreover, with a constant normalization on free complex constants $c_1$ and $c_2$, the wave function $\phi(x,t)$ becomes
\begin{eqnarray} \label{e:phisinh}
&& \phi(x,t,\lambda)= \frac{1}{\sqrt{h-1}} \left(
                    \begin{array}{c}
 \sinh\left[A+ \frac{1}{2}\ln\left(h+\sqrt{h^2-1}\right)\right] \\
 \sinh \left[-A+ \frac{1}{2}\ln\left(h+\sqrt{h^2-1}\right)\right]  \\
                    \end{array}
                  \right). \nonumber\\
                   \label{Eq:A}
\end{eqnarray}
Here, the scaling constant $\sqrt{h-1}$ is introduced so that this wave function does not approach zero in the limit of $h\to 1$ (i.e., $\lambda\to \textmd{i}$). In the view of wave function and Proposition 3.2, this scaling of the wave function  does not affect the solution except for a simple shifting $\frac{1}{2\sqrt{h^2-1}}\ln\left(|\frac{c_{1}}{c_{2}}|\right)$ occurs on the x-direction.

Next, to derive rogue-wave solutions from  above wave functions and adjoint wave functions, we need to choose spectral parameters $\lambda_{k}$ and $\zeta_{k}$ so that the exponents $A$ can vanish under certain limits. Thus, we can take $\lambda_{k}$ to be $\textmd{i}$, and correspondingly,  $\zeta_{k}$ to be $-\textmd{i}$.
   Specifically, we choose spectral parameters $\lambda=\lambda_{k}$, $\zeta=\zeta_{k}$  and  complex parameters $\theta=\theta_{k}$,  where
\begin{eqnarray}
&& \lambda_k=\textmd{i}(1+\epsilon_k^2), \quad  \zeta_k=-\textmd{i}(1+\epsilon_k^2),  \label{e:parametersI} \\
&& \theta_k=\sum_{j=0}^{N-1}s_{j}\epsilon_{k}^{2j},  \quad 1\le k\le N,  \label{e:parametersII}
\end{eqnarray}
$s_0, s_1, \dots, s_{n-1}$ are  $k$-independent  complex constants.

Thus,  with $\lambda=\textmd{i}(1+\epsilon^2),  \zeta=-\textmd{i}(1+\widetilde{\epsilon}^2)$,  we have the following expansions:
\begin{eqnarray}
&&\phi(x,t,\lambda)=\sum_{k=0}^{\infty} \phi^{(k)} \epsilon^{2k},\
\psi(x,t,\zeta)=\sum_{k=0}^{\infty} \psi^{(k)} \widetilde{\epsilon}^{2k},\\
&&\frac{\psi(x,t,\zeta)\phi(x,t,\lambda)}{\lambda-\zeta}=
\sum_{k=0}^{\infty}\sum_{l=0}^{\infty}m_{k,\hspace{0.04cm} l}\hspace{0.04cm}
\widetilde{\epsilon}^{2k} \epsilon^{2l},
\end{eqnarray}
where
\[
\phi^{(k)}=\lim_{\epsilon\rightarrow 0}\frac{\partial^{2k}\phi(x,t,\lambda)}{(2k)! \partial \epsilon^{2k}}, \quad
\psi^{(k)}=\lim_{\epsilon\rightarrow 0}\frac{\partial^{2k}\phi^T(x,-t,\lambda)}{(2k)! \partial \epsilon^{2k}},
\]
\begin{eqnarray}
&&m_{k,l}=\nonumber \\
&&\lim_{\epsilon, \widetilde{\epsilon} \rightarrow 0}\frac{1}{(2k-2)!(2l-2)!}\frac{\partial^{2k+2l-4}}{\partial \widetilde{\epsilon}^{2k-2}\partial \epsilon^{2l-2}}
\left[\frac{\phi^{T}(-t,\zeta)\phi(t,\lambda)}{\lambda-\zeta}\right].  \nonumber\\
\label{detmij1}
\end{eqnarray}

Here, we  use a different  notation $\widetilde{\epsilon}$ instead of $\epsilon$ in the adjoint wave function $\phi$ for above expansions. In the sense of limit $\epsilon\rightarrow 0$, $\widetilde{\epsilon}\rightarrow 0$, they still meets the reduction condition (\ref{PhiPsicond}).  Therefore, applying these expansions to each matrix element in the B\"acklund transformation (\ref{gDTpotential}), performing simple determinant manipulations and taking the limits of $\epsilon_{k},  \widetilde{\epsilon}_{k} \rightarrow 0$ ($1\le k\le N$), we derive general rogue waves in the following  Theorem.

\textbf{Theorem 3.1}.  \emph{The N-th order rogue waves in the focusing  reverse-time nonlocal NLS equation  can be formulated as:}
\[ \label{N-Rws1}
u_N(x,t)=e^{-2 i t} \left(1+2 \textmd{i} \frac{\tau_1}{\tau_0}\right) ,
\]
\emph{where}
\[
\tau_0= \det_{1\leq i,j \leq n}\left(m_{i,j}\right),\ \ \tau _1=\det\left(   \begin{array}{cc}
                        \left(m_{i,j}\right)_{1\leq i,j \leq N} & Y_{(2)} \\
                        X_{(1)} & 0
                      \end{array}
\right),\nonumber
\]
\[
X= \left[\phi^{(0)}, \phi^{(1)},..., \phi^{(N-1)}\right], \
Y=  \left[ \psi^{(0)}, \\ \psi^{(1)} \\ ,...,\\
\psi^{(N-1)} \right]^T.  \nonumber
\]
\emph{Here, $X_{(1)}$ stands the 1-st row in matrix $X$, while $Y_{(2)}$ represents the 2-nd column in   $Y$.  Functions $\phi(\lambda)$ and $\psi(\zeta)$ are defined in the form as (\ref{Adeigen})-(\ref{e:phisinh}), the matrix element $m_{i,j}$ is given through limitation (\ref{detmij1}).}

Since wave functions and adjoint wave functions in the Darboux transformation for the reverse-time nonlocal NLS equation (\ref{e:PTNLS}) are related via the reverse-time reduction, this new reduction will lead to different types of rogue waves,  which  will be discussed in the following.

\subsection{\label{sec:level2}Dynamics in the  rogue-wave solutions}
In this section, we give a discussion on the dynamics of these rogue-wave solutions.
\subsubsection{\label{sec:level3}Fundamental rogue-wave solution}
The first-order (fundamental) rogue wave is obtained by setting $N=1$ in Eq. (\ref{N-Rws1}), where the matrix elements can be obtained from Theorem 3.1. In this case,
\begin{eqnarray*}
&& \tau_{0}=m_{1,1}=-\frac{\textrm{i}}{2}  \left(8 s_0 x+4 s_0^2+16 t^2+4 x^2+1\right), \\
&& \tau_{1}=
\left|
\begin{array}{cc}
 m_{1,1} & Y_{2} \\
 X_{1} & 0 \\
\end{array}
\right|=\frac{1}{2}   \left(8 s_0 x+4 s_0^2+16 t^2+8 \textrm{i} t+4 x^2-1\right),
\end{eqnarray*}
where   $s_0$ is complex constants. Thus, the first-order  rogue-wave is  given by
\begin{eqnarray*}
&& u_1(x,t)=e^{-2 \mathbf{i}\hspace{0.06cm}t} \left(1+2\  \textmd{i} \frac{\tau_1}{\tau_0}\right) \\
&& =-\frac{e^{-2 \mathbf{i} t} \left(8 s_0 x+4 s_0^2+16 t^2+16 \mathbf{i} t+4 x^2-3\right)}{8 s_0 x+4 s_0^2+16 t^2+4 x^2+1}.
\end{eqnarray*}
Denoting $x_{0}=\textrm{Re}\left(s_{0}\right)$ and $y_0=\textrm{Im}\left(s_{0}\right)$, this rogue wave can be rewritten as
\begin{eqnarray}\label{rogueu11}
\hspace{-0.5cm}u_1(x,t)=-e^{-2\mathbf{i}\hspace{0.06cm} t}\left[1+\frac{4(4 \textrm{i} t-1)}{16 t^2+4 \left(\hat{x}+\textrm{i} y_0\right){}^2+1}\right],\ \
\end{eqnarray}
where $\hat{x}=x+x_{0}$. Hence,  this solution has one non-reducible real parameter $y_0$, since the parameter $x_0$ can be removed by space-shifting.  Moreover, it is interesting to find that this  fundamental rogue-wave solution  also admits the first order rogue wave recently derived \cite{JBY2017} in the $\mathcal{PT}$-symmetric nonlocal NLS equation. This is due to an important property which solution (\ref{rogueu11}) satisfies:
\begin{equation}\label{Solutionproperty}
u^*_{1}(-\hat{x},t)=u_{1}(\hat{x},-t)
\end{equation}
For this rogue wave solution, when $y_{0}^2 < 1/4$, it is shown to be nonsingular. If $y_0=0$, it degenerates into the classical  Peregrine soliton for the local NLS equation. [note that any   solution of the local NLS equation satisfying $u^*(x,t)=u(x,-t)$ would satisfy the reverse-time nonlocal equation (\ref{e:PTNLS})]. The peak amplitude of this Peregrine soliton is 3, i.e., three times the level of the constant background. But if $y_0\ne 0$, its peak amplitude would accurately becomes $\left|\frac{4s_{0}^2+3}{4s_{0}^2-1}\right|$,  which is higher than 3.  One of such solutions is shown in Fig. 1.
\begin{figure}[htbp]
   \begin{center}
   \vspace{-1cm}
   \includegraphics[scale=0.220, bb=0 0 385 567]{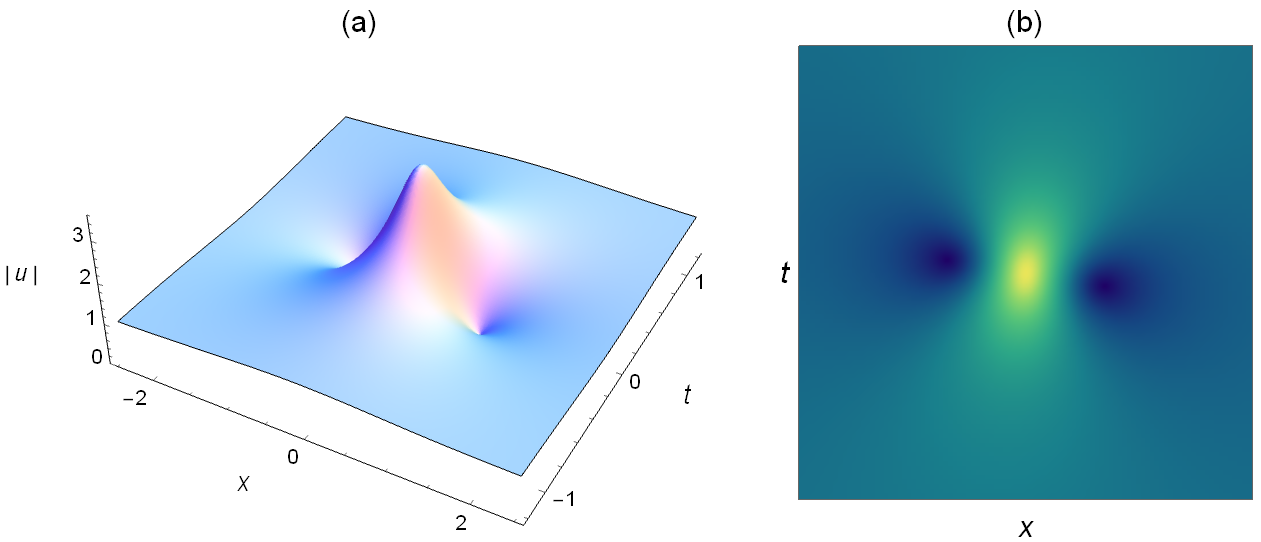} \hspace{5.5cm}
   \caption{The nonsingular first-order rogue wave solution (\ref{rogueu11}). (a). $s_0=r_{0}=\frac{\textrm{i}}{6}$(corresponding to $y_0=1/6$); (b) is the  corresponding density plot.}
   \vspace{-0.25cm}
   \label{Tu1}
   \end{center}
\end{figure}

However, once $y_{0}^2\geq 1/4$, this rogue wave would blow-up at $x=0$ with two time points $t_{c}=\pm \sqrt{(4y_{0}^2-1)/16}$.   One such blowing-up solution is displayed in Fig.2.  Since wave collapse has been reported in bright solitons \cite{AblowitzMussPRL2013,AblowitzMussNonli2016} for the nonlocal NLS equation (\ref{e:PTNLS}).  Here we see that collapse occurs for rogue waves in the reverse-time nonlocal equation as well.
\begin{figure}[htbp]
   \begin{center}
   \includegraphics[scale=0.220, bb=0 0 385 567]{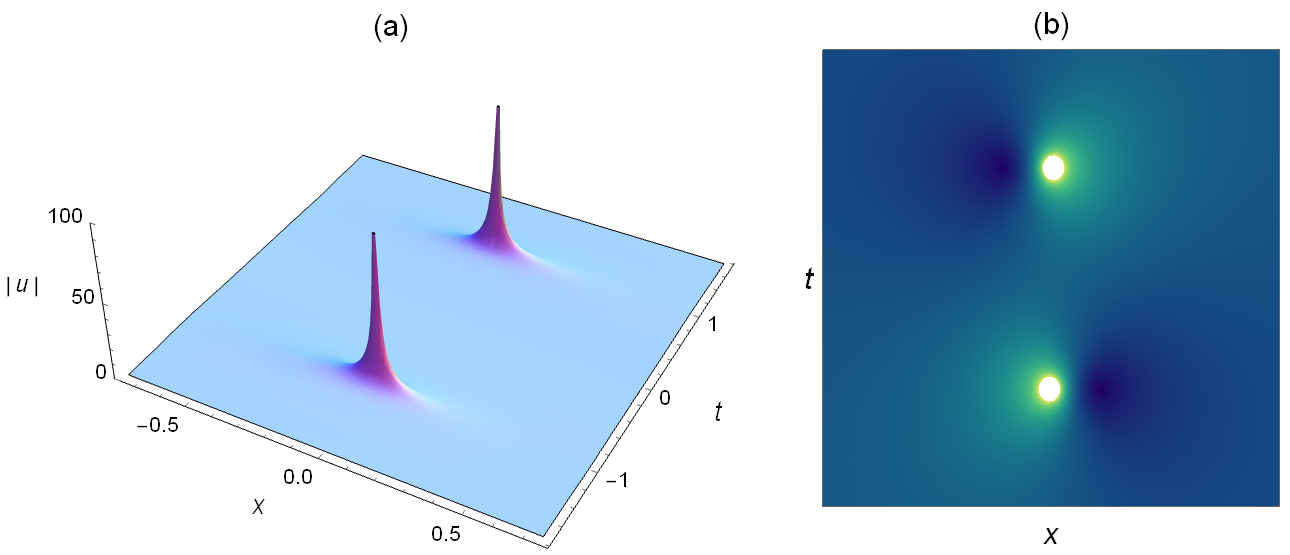} \hspace{5.5cm}
   \caption{A collapsing first-order rogue wave (\ref{rogueu11}) with
   $s_0=2 \textrm{i};$ (a) 3D plot; (b) density plot.}
   \vspace{-0.5cm}
   \label{Tu2}
   \end{center}
\end{figure}
\subsubsection{\label{sec:level3}Higher-order  rogue-wave solution}
Next, we consider the second order  rogue waves, which are given in (\ref{N-Rws1}) with $N=2$.  In this case, the general second order rogue wave can be obtained as
\begin{eqnarray}\label{Second-orderRws}
q_{2}(x,t)=e^{-2 \mathbf{i}\hspace{0.06cm}t} \left(1+2\  \textmd{i} \frac{\tau_1}{\tau_0}\right)
\end{eqnarray}
where,
\begin{widetext}
\begin{eqnarray*}
&&\tau_0=-\frac{4}{9}\{4096 t^6+3072 t^4 x^2+768 t^2 x^4+64 x^6+6912 t^4-1152 t^2 x^2+1584 t^2+48 x^4 \\
&&\hspace{0.75cm}+108 x^2+384 s_0^5 x+48 s_0^4 \left(16 t^2+20 x^2+1\right)+64 s_0^3 \left[x \left(48 t^2+20 x^2+3\right)-3 s_1\right]+64 s_0^6\\
&&\hspace{0.75cm}-48 s_1 x \left(-48 t^2+4 x^2-3\right)+12 s_0^2 \left[-48 s_1 x+256 t^4+96 t^2 \left(4 x^2-1\right)+80 x^4+24 x^2+9\right]\\
&& \hspace{0.75cm} +144 s_1^2+24 s_0 \left[6 s_1 \left(16 t^2-4 x^2+1\right)+x \left(256 t^4+128 t^2  x^2+16 x^4+8 x^2-96t^2+9\right)\right]+9\},\\
&&\tau_1=-\frac{8}{3}\left[1024 t^5 + \textrm{i}  \left(1280  t^4+16 x^4+64  s_0 x^3+24  x^2+96  s_0^2 x^2+48  s_0 x+64 s_0^3 x+48 s_1 x\right) \right.\\
&& \hspace{-0.0cm} + \textrm{i} t^2  \left(384  x^2+768  s_0 x+384 s_0^2+288 \right)+t^3 \left(512 x^2+1024 s_0 x+512 s_0^2+128\right)+t\left(256 s_0 x^3+384 s_0^2 x^2\right.\\
&&\left. \left. \hspace{-0.0cm} +256 s_0^3 x-192 s_0 x+192 s_1 x+64 s_0^4-96 s_0^2+192 s_0 s_1+64 x^4-96 x^2-60\right)+\textrm{i} \left(24 s_0^2+16 s_0^4+48 s_0 s_1-3\right)\right],
\end{eqnarray*}
 \end{widetext}
 
By choosing free complex parameters $s_0 $ and $s_1$ in (\ref{Second-orderRws}), we can get both nonsingular and singular (blowing-up) solutions. One  nonsingular  triangular pattern  solution  is observed  and displayed in Fig. 3(a). Even though the present solution does not admit $u^*(x,t)=u(x,-t)$  and thus does not satisfy the local NLS equation,  this pattern resemble that in the local NLS equation \cite{AAS2009,DGKM2010,DPMB2011,GLML2012,OhtaJY2012}. Moreover, this triangular pattern  features the double temporal bumps with a single temporal bump,  which  is also different from the pattern found in the $\mathcal{PT}$-symmetric nonlocal NLS equation\cite{JBY2017}.
\begin{figure}[htbp]
   \begin{center}
    \vspace{2.5cm}
    \hspace{-0.5cm} \includegraphics[scale=0.190, bb=0 0 385 567]{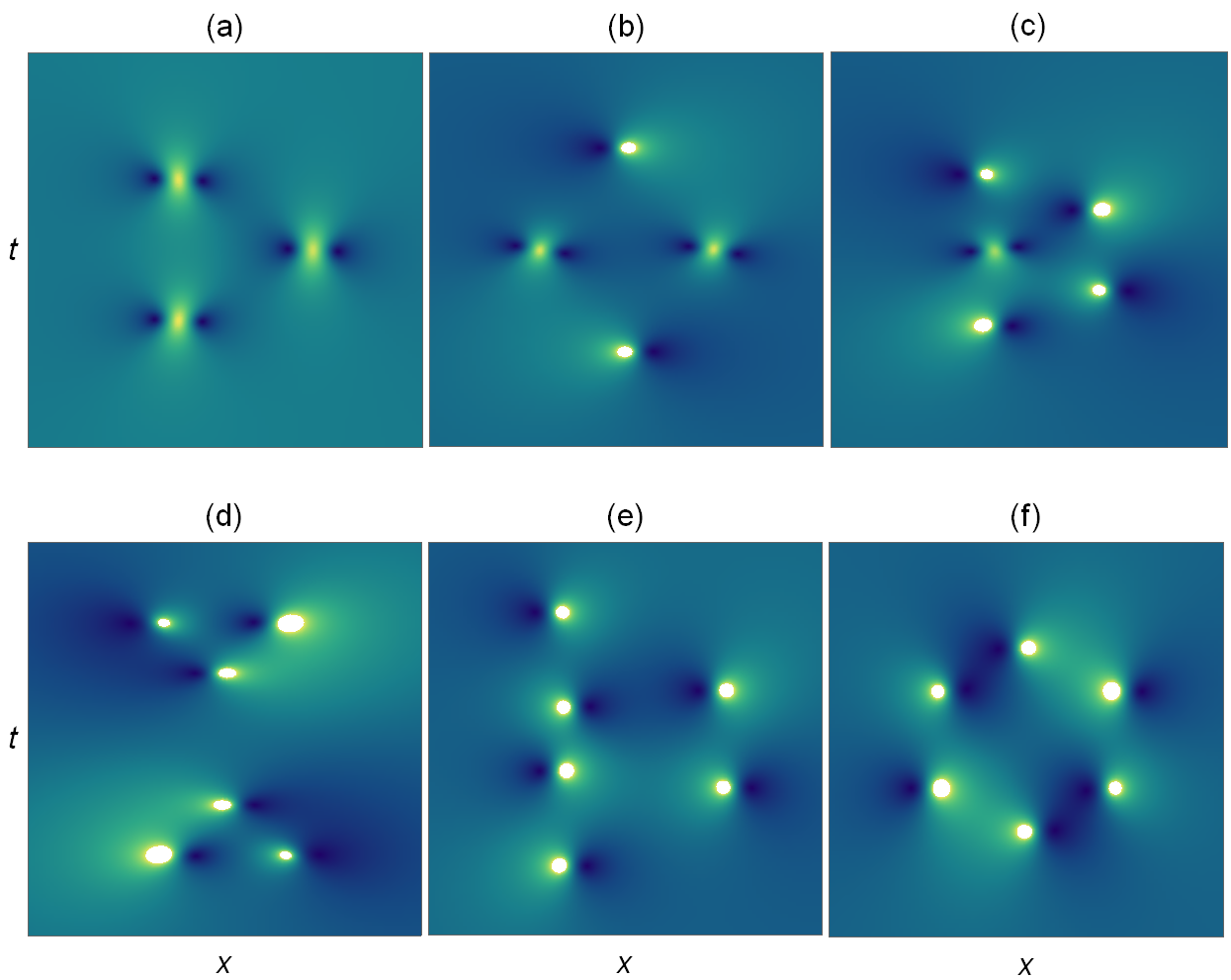} \hspace{5.5cm}
   \caption{Six  second-order rogue waves (\ref{Second-orderRws}) with different patterns.
   (a). $s_0=\frac{i}{20};s_1=20;s_2=0;$
(b). $s_0=2 i;s_1=30 i;$
(c).  $s_0=2 i;s_1=10;$
(d). $s_0=5 i;s_1=0;s_2=0;$
(e). $s_0=2 i;s_1=40;s_2=0;$
(f). $s_0=0;s_1=30 i;s_2=0;$ }
\vspace{-0.15cm}
   \label{Tu2}
   \end{center}
\end{figure}
For the second-order rogue waves,  the blowing-up solutions are shown to have more interesting and complex patterns,   which  have not been observed before. Five of them are displayed in Fig. 3. In panel (b), the solution contains two singular (blowing-up) peaks on the horizontal $x$ axis, and two nonsingular ``Peregrine-like"  humps on the vertical $t$ axis. In panel (c), these is one ``Peregrine-like" nonsingular hump in the middle of two singular peaks on the vertical $t$ axis, plus another two singular peaks locate also on the vertical $t$ axis.  The other three panels each contain six singular peaks, which are arranged in different patterns. The maximum number of singular peaks in these solutions is six.

\begin{figure}[htbp]
   \begin{center}
   \vspace{1.5cm}
    \hspace{-2.5cm} \includegraphics[scale=0.260, bb=0 0 385 567]{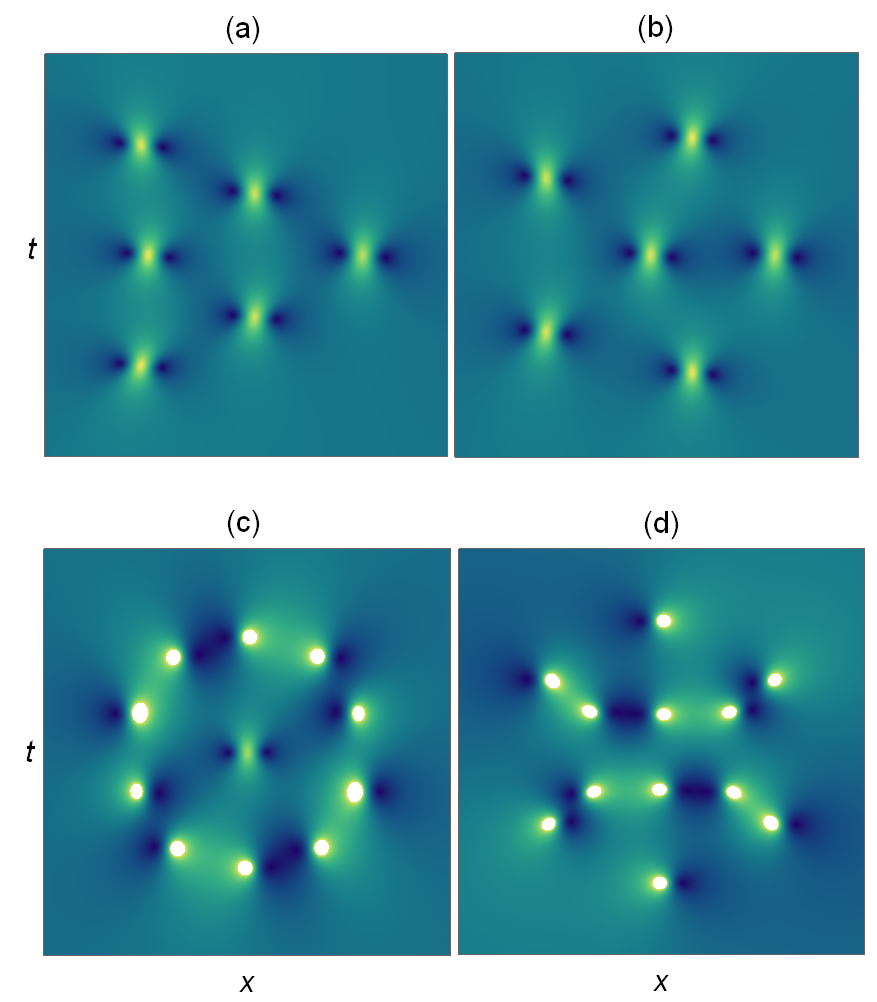}
   \caption{ Four third-order rogue waves (top row: nonsingular solutions; lower row: collapsing solutions).
   (a).  $s_0=\frac{i}{10};s_1=20;s_2=0$
     (b). $s_0=\frac{i}{10};s_1=0;s_2=-300;$
     (c).  $s_0=0;s_1=0;s_2=300 i$;  $s_0=4 i;s_1=84;s_2=0;$
     (d). $s_0=2i;s_1=0;s_2=300 i$; }
     \vspace{-0.5cm}
   \label{Tu2}
   \end{center}
\end{figure}
Third-order rogue waves would exhibit an even wider variety of solution patterns. Here,  four of them are chosen and displayed in Fig. 4. The upper row shows two nonsingular solutions, which contain six ``Peregrine-like" humps arranged in triangular and pentagon patterns,  resemble those solutions in the local NLS equation \cite{GLML2012,OhtaJY2012}, but not in the  $\mathcal{PT}$-symmetric nonlocal NLS equation\cite{JBY2017},  while the spatial-temporal structures displayed there are  different from these two nonsingular patterns.

The lower row in Fig. 4 contains  two blowing-up solutions.   In panel (c), there is  one ``Peregrine-like" nonsingular hump surrounded by ten singular peaks. In panel (d), twelve singular peaks arranging in a very exotic pattern. Again, this maximum number of singular peaks is found to be twelve.

The results given above can be apparently extended to higher order rogue waves. By special choices of the free parameters $s_{k}\ (k \in \mathds{N}^{+})$, we
can yield even richer spatial-temporal patterns, in the form of nonsingular humps,  singular peaks or their hybrid  patterns,  but with more intensity.
\begin{figure}[htbp]
   \begin{center}
   \vspace{0.5cm}
   \hspace{-0.5cm}\includegraphics[scale=0.255, bb=0 0 385 567]{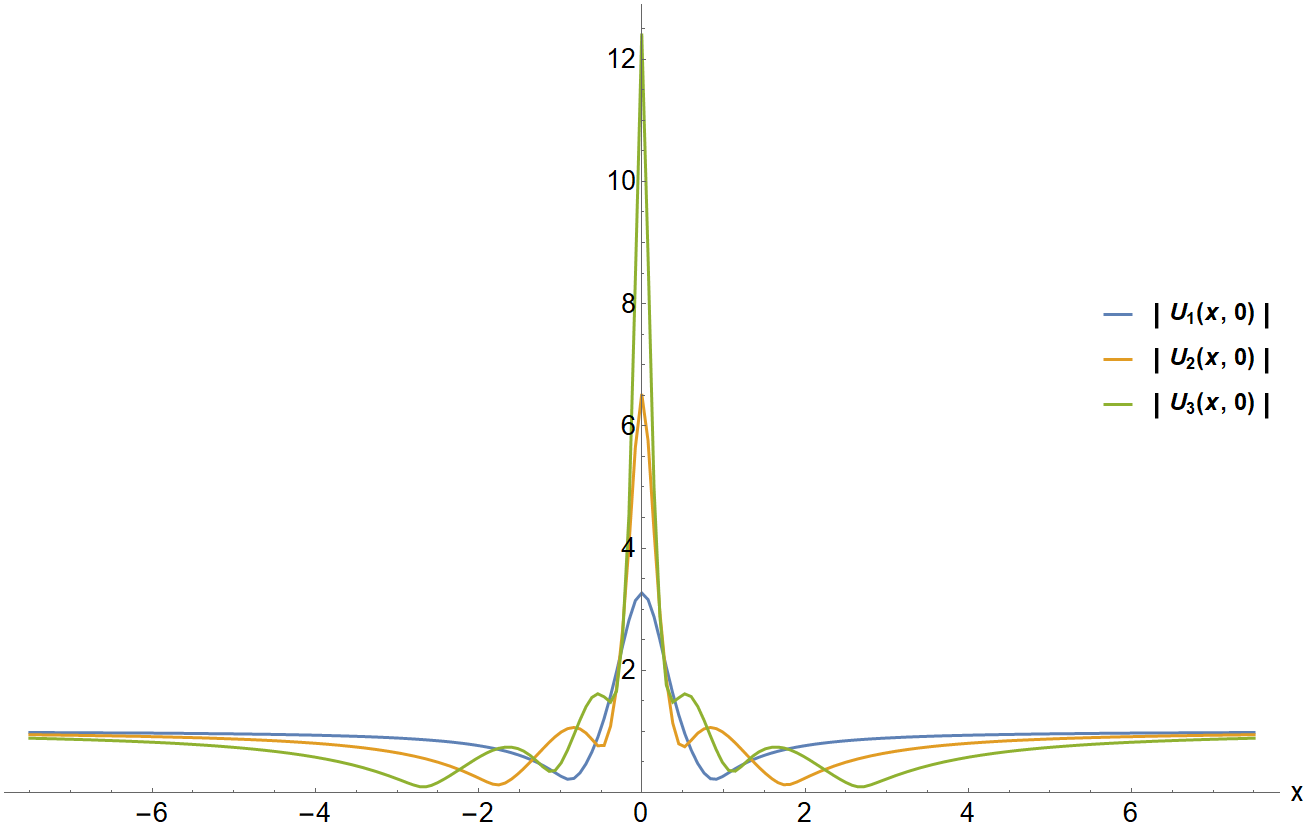} \hspace{4.5cm}
   \caption{The 2D plot of $|u_{N}(x,0)|$ from the first- to the  third-order rogue waves, with parameters:
   $s_0=\frac{i}{8};s_1=0;s_2=0;$  }
   \vspace{-0.5cm}
   \label{Tu2}
   \end{center}
\end{figure}
Moreover,  for the fundamental pattern rogue waves(i.e., only $s_{0}$ are taken as nonzero value while the rest of the parameters are taken as zero.) in the local NLS equation,  the central maximum amplitude  for each  solutions is given by $|u_{N}(0,0)|=2N+1$.  However, this value becomes higher for nonlocal equation (\ref{e:RTNLS}). Taking  the first to the third order fundamental rogue-wave pattern as an example.  It is  shown that the elevations for them are  numerically attained at  $|u_{1}(0,0)|=3.26667>3$,  $|u_{2}(0,0)|=6.52338>5$,   and $|u_{3}(0,0)|=12.4318>7$ (See Fig.5).  Thus, such rogue waves can be fairly dangerous if they arise in physical situations.
\section{Rogue waves in the reverse-time nonlocal DS systems}
In this part, we consider rogue waves in the reverse-time nonlocal DS equations, which is recently introduced in refs.\cite{AblowitzMussSAPM}.  Eqs.(\ref{PTDS})-(\ref{PTDSS}) can be regarded  as an integrable multidimensional version of the reverse-time nonlocal  nonlinear Schr\"{o}dinger equation.

We start with the following  auxiliary linear system:
\begin{eqnarray}
&&\emph{L}\Phi=0, \emph{L}=\partial_{y}-J\partial_{x}-P, \label{Lax-pair1} \\
&& \emph{M}\Phi=0, \emph{M}=\partial_{t}-\sum_{j=0}^{2}V_{2-j}\partial^{j},\label{Lax-pair2}\\
&& V_{0}=\textrm{i} \gamma^{-1} J,\ V_{1}= \textrm{i}  \gamma^{-1} P, \ V_{2}=\frac{\textrm{i}}{2\gamma}\left(P_{x}+\gamma^2 J P_{y}+Q \right),\nonumber \\
 && J=\gamma^{-1}\left(
                    \begin{array}{cc}
                      1 & 0\\
                      0 & -1 \\
                    \end{array}
                  \right),\
  P=              \left(
                    \begin{array}{cc}
                      0 & q\\
                      -r & 0 \\
                    \end{array}
                  \right),\
  Q=
\left(
\begin{array}{cc}
\phi_{1} & 0\\
0 & \phi_{2}\\
\end{array}
\right),   \nonumber
\end{eqnarray}
with,
\begin{eqnarray}
\phi=qr-\frac{1}{2 \gamma}(\phi_{1}-\phi_{2}).
\end{eqnarray}
The compatibility condition  $\left[ L, M \right]=0$ leads to the following equation:
\begin{equation}\label{zerocurvatureeq}
  P_{t}=V_{2,y}-J V_{2,x}-\left[  P, V_{2} \right] + \sum_{j=1}^{2}V_{2-j}\partial^{j}(P),
\end{equation}
which yields to the following coupled system for potential functions $q(x,y,t)$, $r(x,y,t)$  and $\phi(x,y,t)$:
\begin{eqnarray}
  && iq_{t}+\frac{1}{2}\gamma^{2} q_{xx}+\frac{1}{2} q_{yy}+ (q r - \phi)q=0, \label{GePTDS} \\
  && \phi_{xx}-\gamma^{2} \phi_{yy} - 2\left(q r\right)_{xx}=0.\label{GePTDSS}
\end{eqnarray}
Then, the  reverse-time nonlocal DS equations (\ref{PTDS})-(\ref{PTDSS}) are obtained from the above coupled system (\ref{GePTDS})-(\ref{GePTDSS}) under the symmetry reduction:
\[ \label{qrphiS}
r(x,y,t)=\sigma q(x,y,-t).
\]
Under reduction (\ref{qrphiS}), it is noticed that  potential matrix  $P$ satisfies the following  symmetry condition:
\begin{eqnarray}\label{R-TsymmetryP}
\hspace{-0.5cm} \tau_{\sigma}  \overline{P}(x,y,t)  \tau_{\sigma}^{-1}=-P^{\dag}(x,y,-t), \ \ \tau_{\sigma}=\left(
                                                            \begin{array}{cc}
                                                              1 & 0 \\
                                                              0 & \sigma \\
                                                            \end{array}
                                                          \right).
\end{eqnarray}
Here $\bar{}$ represents the complex conjugation of the function.   In this case, we should have $\phi_{1}(x,y,t)=\phi_{1}(x,y,-t),\  \phi_{2}(x,y,t)=\phi_{2}(x,y,-t)$.  Thus, $\phi(x,y,t)$ is an even-function on $t$,  and $V_{2}(x,y,t)$ have the property
\begin{equation}\label{R-TsymmetryV2}
\tau_{\sigma}  \overline{V}_{2}(x,y,t)\tau_{\sigma}^{-1} = V_{2}^{\dag}(x,y,-t)+\textrm{i}\gamma P_{x}^{\dag}(x,y,-t).
\end{equation}
Afterwards, symmetry conditions (\ref{R-TsymmetryP})-(\ref{R-TsymmetryV2}) further lead to the symmetry in $L$ and $M$:
\begin{eqnarray}\label{R-TsymmetryLM}
  \tau_{\sigma} L^{\dag} \tau_{\sigma}^{-1}=-\overline{L}_{\left(t \rightarrow -t\right)},\   \  \tau_{\sigma} M^{\dag} \tau_{\sigma}^{-1}=\overline{M}_{\left(t\rightarrow -t \right)}.
\end{eqnarray}
Here, $\dag$ stands the (formal) adjoint on an operator.  Especially, for a matrix $A$, where $A^{\dag}$ denotes the Hermitian conjugate of $A$.

\subsection{\label{sec:level2} A  unified  binary Darboux transformstion and  its reductions}
In this section, we construct the corresponding Darboux transformation in a unified way for the reverse-time nonlocal DSI  and DSII equation. It is known that the DT for DSI and DSII equations (local or nonlocal) appears either in the differential form or in the integral form, which are totally different. However,  for this reverse-time nonlocal DS system, we can find a unified way to tackle their DTs and represent the solutions in one formula.

The standard scheme for  binary DT was firstly introduced by Matveev  and Salle.\cite{MatveevDT1991} Especially, for operators $L$  and $M$ given in (\ref{Lax-pair1})-(\ref{Lax-pair2}), a  binary DT has been constructed,\cite{MatveevDT1991, NimmoGY2000} which can be written explicitly as
\begin{eqnarray}\label{BinaryDT}
\hspace{-0.5 cm}G_{\theta, \rho}=I -\theta \Omega^{-1}(\theta,\rho)\partial^{-1}\rho^{\dag}, \ \ \Omega(\theta,\rho)=\partial^{-1} ( \rho^{\dag} \theta ),
\end{eqnarray}
where $\theta$  satisfy  $L(\theta)=0$, and  $\rho$ admits the adjoint operator: $L^{\dag}(\rho)=0$. Then the operator
\begin{eqnarray}
 \hat{L}=G_{\theta, \rho}LG_{\theta, \rho}^{-1}
\end{eqnarray}
can be directly verified to be a new operator which has the same form as $L$, and so is for $M \rightarrow \widehat{M}$. Furthermore, this Darboux transformation makes sense for any $m\times k$ matrices $\theta$ and $\rho$, and we only need   $\Omega(\theta,\rho)$ to be an invertible square matrix \cite{NimmoGY2000}.  By the  combination of an elementary DT with its inverse, one derives the  B\"{a}cklund transformation between  the potential matrices:
\begin{eqnarray}\label{BinaryDTpotantial}
 \hat{P}=P+ [J, \theta \Omega^{-1}(\theta,\rho)\rho^{\dag}].
 \end{eqnarray}
Next, to reduce (\ref{BinaryDT})  and (\ref{BinaryDTpotantial})  to the Binary DT for the   reverse time nonlocal DS equations (\ref{PTDS})-(\ref{PTDSS}),   the wave function  and the adjoint wave function  are restricted  to satisfy the following important relation:
\begin{eqnarray}\label{conditionDS-IBDT}
\rho(x,y,t)=\tau_{\sigma} \overline{\theta}(x,y,-t).
\end{eqnarray}

With   condition (\ref{conditionDS-IBDT}), it can be verified that the new potential matrix in (\ref{BinaryDTpotantial}) indeed preserves the symmetry:
\begin{eqnarray}\label{R-TsymmetrynewP}
\tau_{\sigma}  \overline{\hat{P}}(x,y,t)\tau_{\sigma}^{-1}=-\hat{P}^{\dag} (x,y,-t).
\end{eqnarray}
At the same time,  $\hat{V}_{2}$  will also preserves symmetry (\ref{R-TsymmetryV2}) due to (\ref{R-TsymmetrynewP}). Hence new operators $\hat{L}$ and $\hat{M}$ can further satisfy symmetry (\ref{R-TsymmetryLM}). Therefore, under this reduction, matrix (\ref{BinaryDT})  becomes the binary DT for the  reverse-time nonlocal DS equations. The B\"{a}cklund transformations (\ref{BinaryDTpotantial}) for generating  new solutions are given as
\begin{eqnarray}
&&  \hat{q}=q+  \gamma^{-1}[\theta \Omega^{-1}(\theta,\rho)\rho^{\dag} ]_{1,2}, \label{BacklundBDTDSII} \\
&&  \hat{\phi}=\phi + 2 \gamma^{-2} [\textbf{tr}(\theta \Omega^{-1}(\theta,\rho)\rho^{\dag})]_{x}. \label{BacklundBDTDSIIw}
\end{eqnarray}

Taking $n$ different wave-functions $\theta_{i}$ and adjoint wave-functions $\rho_{i}$ which satisfy  reduction (\ref{conditionDS-IBDT}). Denoting
\begin{eqnarray*}
\Theta=\left( \theta_{1},\cdots, \theta_{n} \right),\ \textbf{P}=\left( \rho_{1},\cdots, \rho_{n} \right).
\end{eqnarray*}
Introducing $1\times n$ row vectors $\psi_{i}$ and $\varphi_{i}$, $i=1,2$. The above $2\times n$ matrices can be further represented by
\begin{eqnarray*}
   \Theta=\left(
            \begin{array}{c}
              \psi_{1} \\
               \psi_{2} \\
            \end{array}
          \right),\ \ \
           \textbf{P}=\left(
            \begin{array}{c}
              \varphi_{1} \\
               \varphi_{2} \\
            \end{array}
          \right).
\end{eqnarray*}
Defining the $n \times n$ matrix $\Omega(\Theta,\textbf{P})$  as:
\begin{eqnarray*}
\Omega(\Theta,\textbf{P})=\left( \partial^{-1} ( \rho_{i}^{\dag} \theta_{j} )\right)_{1\leq i,j \leq n}.
\end{eqnarray*}
Hence, the n-fold Buckl\"{u}nd transformations  between  potential functions is given as:
\begin{eqnarray}
&&q_{[n]}(x,y,t)= u_{[0]}(x,y,t)-\frac{2}{\gamma}  \frac{  \left|\begin{array}{cc}
 \Omega(\Theta,\textbf{P}) & \varphi_{2}^{\dag}  \\
 \psi_{1} & 0
\end{array}\right| }{\det(\Omega(\Theta,\textbf{P}))}, \label{un+1} \\
&& \phi_{[n]}(x,y,t)=  w_{[0]}(x,y,t) + 2 \gamma^2 \partial_{x}^{2}\{\log \left[\det\left( \Omega(\Theta,\textbf{P})\right) \right] \}. \nonumber \\
&& \label{wn+1}
\end{eqnarray}
This n-fold Buckl\"{u}nd transformation  has been reported \cite{BYnonlocalDS} by using quasi-determinant technique with simple linear algebra.  The  proof of  expressions (\ref{un+1})-(\ref{wn+1}) has already been given \cite{BYnonlocalDS}.
\subsection{General binary Darboux transformation}

To obtain the high-order solutions, we need to perform the general DT scheme.  The idea was firstly introduced by Matveev\cite{MatveevDT1991} and  further developed by Ling \cite{GLML2012} to construct  high-order rogue waves in local NLS equation. For this nonlocal DS system, the general binary DT can be also constructed via a direct limitation technique, those  are summarised as the following results.

\emph{\textbf{Theorem 3}}
\emph{The generalized binary Darboux  matrix for the reverse time nonlocal DS equations (\ref{PTDS})-(\ref{PTDSS}) can be represented as:}
\begin{eqnarray*}
G_{[n]}=I- \Theta  \Omega^{-1}(\Theta,\textbf{P})\Omega(\cdot ,\ \textbf{P}), \label{gDTRTDS}
\end{eqnarray*}
\emph{where,}
\begin{eqnarray*}
&&\theta_{i}=\theta_{i}(k_{i}+\epsilon_{i}),\ \rho_{j}=\rho_{j}(\tilde{k}_{j}+\tilde{\epsilon}_{j}),\\
 && \widehat{\Theta}=\left( \Theta_{1}, \Theta_{2},\ldots, \Theta_{s} \right),\  \Theta_{i}=\left(\theta_{i}^{[0]},\ldots,\theta_{i}^{[r_{i}-1]}\right),\\
 && \widehat{\textbf{P}}=\left( \textbf{P}_{1}, \textbf{P}_{2},\ldots, \textbf{P}_{s} \right),\ \textbf{P}_{j}=\left(\rho_{j}^{[0]}, \ldots, \rho_{j}^{[r_{j}-1]}\right), \\
 &&  \Omega(\widehat{\Theta},\widehat{\textbf{P}})=\left( \Omega^{[i j]}\right)_{1\leq i,j \leq s}, \ \Omega^{[ij]}=\left(\Omega^{[ij]}_{m,n} \right)_{r_{i}\times r_{j}}, \\
 && \theta_{i}=\sum_{k=0}^{r_{i}-1}\theta_{i}^{[k]}\epsilon_{i}^{k}+\mathcal{O}(\epsilon_{i}^{r_{i}}),\  \rho_{j}=\sum_{k=0}^{r_{j}-1}\rho_{j}^{[k]}\tilde{\epsilon}_{j}^{k}+\mathcal{O}(\tilde{\epsilon}_{j}^{r_{j}}),\\
 &&\Omega(\theta_{j}(k_{j}+\epsilon_{j}), \rho_{i}(\tilde{k}_{i}+\tilde{\epsilon}_{i}))=\sum_{m=1}^{r_{i}}\sum_{m=1}^{r_{i}} \Omega^{[ij]}_{m,n}+\mathcal{O}(\tilde{\epsilon}_{i}^{r_{i}},\epsilon_{j}^{r_{j}}).
\end{eqnarray*}
\emph{Moreover, the general  Buckl\"{u}nd transformations between potential functions are given as:}
\begin{eqnarray}
&& q_{[n]}=q_{[0]}- \frac{2}{\gamma}  \frac{  \left|\begin{array}{cc}
 \Omega(\widehat{\Theta},\widehat{\textbf{P}}) & \widehat{\textbf{P}}_{2}^{\dag}  \\
 \widehat{\Theta}_{1} & 0
\end{array}\right| }{\det(\Omega (\widehat{\Theta},\widehat{\textbf{P}}))}, \label{gDTDS-IIQN}\\
&& \phi_{[n]}= \phi_{[0]}- 2\partial_{x}^{2}\{\log \left[\det\left(  \widehat{\Omega}(\Theta,\textbf{P})\right) \right] \},\label{gDTDS-IIwN}
\end{eqnarray}
\emph{Here,} $\widehat{\Theta}_{1}$ \emph{represents the 1-st row of matrix} $\widehat{\Theta}$, \emph{and} $\widehat{\textbf{P}}_{2}$ \emph{stands the 2-nd row in matrix } $\widehat{\textbf{P}}$.

The proof for above results can be obtained  by directly applying the expansions to each matrix element in the n-fold Buckl\"{u}nd transformation, performing simple determinant operations and taking the limits in (\ref{un+1})-(\ref{wn+1}).

\subsection{\label{sec:level2} Rogue-wave solutions  and their dynamics}
It is shown\cite{OhtaJKY2012,OhtaJKY2013} that  a family of  rational solutions were derived by the bilinear method,  which generate the rogue waves for the local DS equations. In this work,  rogue wave solutions for the reverse time nonlocal DS equations can be  constructed via a unified binary Darboux transformation method.

Firstly, choosing a  constant  $q_{[0]}=c,\ \phi_{[0]}=\sigma c^2\ (c \in \mathds{R})$   as the seeding solution.
Then  the eigenfunction  solved from  Eqs. (\ref{Lax-pair1})-(\ref{Lax-pair2})  has the form
\begin{eqnarray*}
 &&\xi_{i}(x,y,t)=c_i \exp \left[\omega _i(x,y,t)\right],\\
  &&\eta_{i}(x,y,t)=\frac{ \lambda_i c_i }{c }\exp \left[\omega _i(x,y,t)\right],\\
  && \omega _i(x,y,t) =\alpha_{i} x+ \beta_{i} y+\gamma_{i} t,  \ \gamma_{i}= \textrm{i}  \gamma^{-1}  \alpha_{i}  \beta_{i}, \\
&& \alpha_{i}=-\frac{1}{2} \gamma \left(\lambda_{i} +\frac{\sigma c^2 }{\lambda_{i} }\right),\   \beta_{i}=\frac{1}{2} \left(\lambda_{i} -\frac{\sigma c^2 }{\lambda_{i} }\right),
\end{eqnarray*}
where  $\lambda_{i}=r_i \exp \left(\textrm{i} \varphi _i \right)$, $r_i$   and $\varphi _i$ are free real parameters, $c_{i}$ is set to be  complex.

Generally, to derive rational type solutions, we choose a more general eigenfunction via superposition principle, which can be written   in the form as:
\begin{eqnarray}\label{Eigenfunctions}
\theta(x,y,t) :=
\{f_k +\partial_{ \varphi_{k}}\}
\left(
 \xi _k,
 \eta _k
\right)^T, \ f_{k}\in \mathbb{C}.
\end{eqnarray}

\subsubsection{ Fundamental rogue waves}
 To derive the first order rational solution,  we set $n=1$, $c=1$ with $c_{1}=1$ in formula (\ref{un+1})-(\ref{wn+1}). Then the  first-order rational solution in this reverse-time nonlocal system is:
\begin{eqnarray}
&&q_{1}(x,y,t)=1-\frac{2 \emph{F}_{1}(t)}{\emph{F}(x,y,t)},  \label{1-stRwu}\\
&&\phi_{1}(x,y,t)=\sigma +2 \gamma^2  [\ln(F(x,y,t))]_{xx}, \label{1-stRww}
\end{eqnarray}
where,
\begin{eqnarray}
 \nonumber
&&\emph{F}_{1}(t)=4 i \gamma^2 (\lambda_{1}^{-2} + \lambda_{1}^{2})t+ 2,\  \lambda_{1}=r_1 e^{\textrm{i} \varphi _1},\\
 \nonumber
&&\emph{F}(x,y,t)= \left[\sigma \lambda_{1}^{-1}(\gamma x+y)  - \lambda_{1} (\gamma x-y)-2\ \textrm{i} f_{1}+1 \right]^2  +  4\left(  \lambda_{1}^{-2}+ \lambda_{1}^{2} \right)^2 t^2 + 1. \nonumber
\end{eqnarray}

The structure of this rational solution is quite different from that in the partially $\cal{PT}$-symmetric nonlocal Davey-Stewartson equations.
By analysing the denominator in solution (\ref{1-stRwu}), it is shown that this  rational solution has different dynamical  patterns according to the parameter values of $ r_{1}$ and $\varphi_{1}$. Those results are discussed separately  in the following context.

\vspace{0.1cm}
 Reverse-time nonlocal DSI ($\gamma^2=1$) equation.
\vspace{0.1cm}

(i). When $\varphi_{1}=k\pi$,  $\lambda_{1}=(-1)^k r_{1}$ is a real number. In this case,  $q_{1}(x,y,t)$ is a line  rogue wave which
approaches a constant background, i.e., $q_{1}\rightarrow 1$, $\phi_{1}\rightarrow \sigma$ as $t\rightarrow \pm \infty$. Defining $ m_{1}=r_{1}-\sigma r_{1}^{-1}, m_{2}=r_{1}+\sigma r_{1}^{-1}$, then $F_{1}(t)$  and $F(x,y,t)$ becomes:
\begin{eqnarray*}
&& \emph{F}_{1}(t)=  2 i (m_{1}^{2} + m_{2}^{2})t+ 2, \\
&& \emph{F}(x,y,t)=\left[ H(x,y)\right]^2+\left(  m_{1}^{2}+ m_{2}^{2} \right)^2 t^2 + 1,\\
&& H(x,y)= m_{1}\gamma x -  m_{2} y + (-1)^k(2 \textrm{i} f_{1}-1).
\end{eqnarray*}
For this solution,  if Re$(f_{1})\neq 0$  and (Re$(f_{1}))^2\geq \frac{1}{4}$, then  function $F(x,y,t)$ becomes zero at symmetrical critical time $t_{c}=\pm \frac{\sqrt{4(\textrm{Re}(f_{1}))^2-1}}{ m_{1}^{2}+ m_{2}^{2}}$,  the singularity  occurs along on the line:
\begin{equation*}
m_{1}\gamma x -  m_{2} y - (-1)^k(2 \textrm{Im}(f_{1})+1)=0
\end{equation*}
in the $(x,y)$ plane.
\begin{figure}[htb]
   \begin{center}
   \vspace{1.5cm}
   \includegraphics[scale=0.200, bb=0 0 385 567]{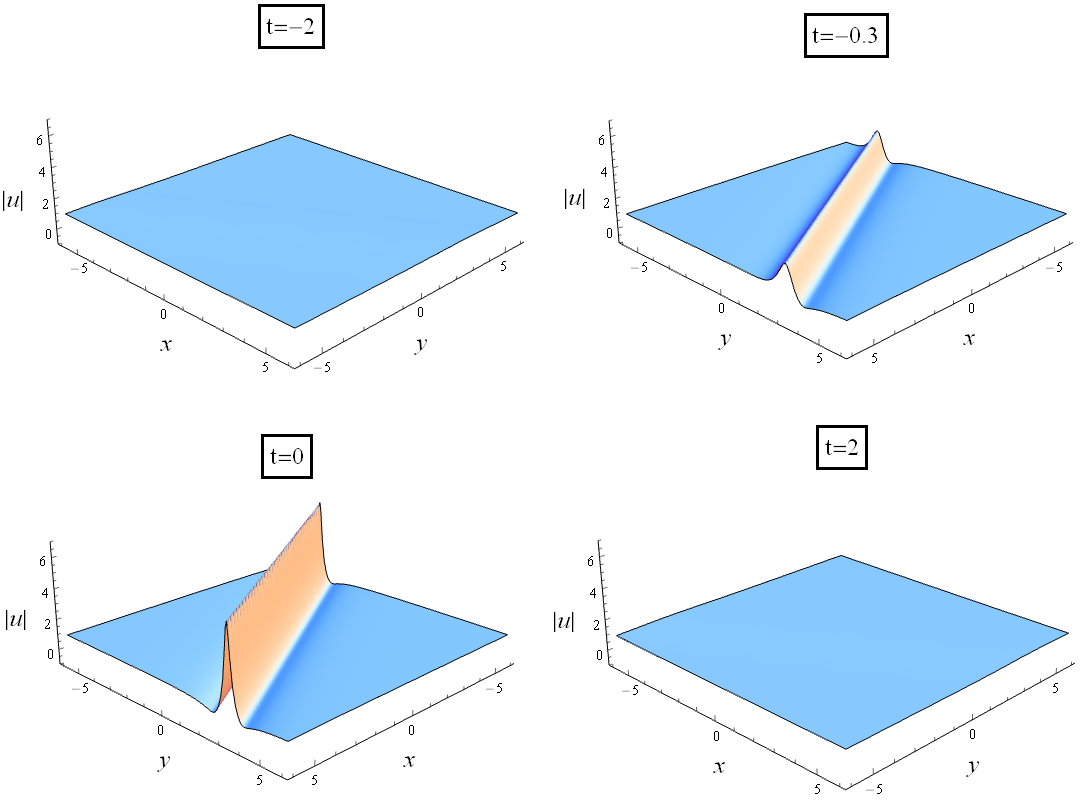} \hspace{5.5cm}
   \caption{Fundamental rogue waves in reverse-time nonlocal DS-I equation  with parameters:
   $k=2$, $\sigma=1$, $r_{1}=2$, $f_{1}=\frac{1}{3}$. }
   \vspace{-0.5cm}
   \label{Tu6}
   \end{center}
\end{figure}

For other choice in the parameter $f_{1}$, this solution describes the nonsingular line rogue waves, The imaginary part of $f_{1}$ can can be further shifted by  $x$  or $y$.  The line in this solution oriented in the $\left(m_{2},  \gamma m_{1} \right)$ direction of the $(x,y)$ plane.  The orientation angle $\beta$ of this solution  is
$\beta=\gamma \arctan(m_{1}/m_{2}),\ (\gamma=\pm 1)$. The width for this line wave is $\sqrt{m_{1}^2+m_{2}^2}= \frac{2}{\sqrt{\sigma \cos 2\beta}}$, so it is angel dependent. Moreover, if  (Re$(f_{1}))^2 < \frac{1}{4}$,  the solution keeps as constant along the line direction with $m_{1}\gamma x -  m_{2} y - (-1)^k(2 \textrm{Im}(f_{1})+1)=0$ fixed.  As $t\rightarrow \pm\infty$,
the solution $q$ uniformly approaches the constant background $1$ everywhere in the spatial plane.  But in the intermediate times, $|q|$ reaches
maximum amplitude  $\frac{4 (\textrm{Re}f_{1})^2+3}{4 (\textrm{Re}f_{1})^2-1}$ (more than three times the background amplitude) at the center  of the line wave at time $t=0$. The speed at which this line approaches its peak amplitude is $\frac{2\sigma}{\cos2\beta}$, which is also angle dependent. Moreover, as what has been discussed in \cite{OhtaJKY2012}, for a given $\sigma$, the line rogue waves in the reverse-time nonlocal DSI equation have a limited range of orientations. When  $\sigma=1$, since $r_{1}$ is a real number,  one can see that $|m_{1}/m_{2}|<1$, thus the orientation
angle of this line rogue wave is between $-45^{\circ}$ and $45^{\circ}$.  If $\sigma=-1$, the satiation is opposite and the  orientation
angle is between $45^{\circ}$ and $135^{\circ}$.
We plot this fundamental line rogue wave in Fig.6 with parameters taken as $k=2$,  $\sigma=1$, $r_{1}=2$ and $f_{1}=\frac{1}{3}$.

Especially, when  $\textrm{Re}(f_{1})=0$, this solution has  the property:
\begin{eqnarray}\label{Con-RTsymmetry}
  \overline{q_{1}}(x,y,t)=q_{1}(x,y,-t),
\end{eqnarray}
which indicates that $q_{1}(x,y,t)$ also admits the local DSI equation.

(ii). When $\varphi_{1}=(\frac{2k-1}{2})\pi$,  then $\lambda_{1}= \textrm{i}(-1)^{k-1} r_{1} $ is a purely imaginary number. This generates the rational travelling waves with:
\begin{eqnarray*}
&& \emph{F}_{1}(t)=  -2 i (m_{1}^{2} + m_{2}^{2})t+ 2, \\
&& \emph{F}(x,y,t)=\left(  m_{1}^{2}+ m_{2}^{2} \right)^2 t^2-\left[ G(x,y)\right]^2 + 1,\\
&& G(x,y)= m_{2}\gamma x -  m_{1} y + (-1)^{k-1}(2f_{1}+\textrm{i}).
\end{eqnarray*}
The ¡°ridge¡± of the solution lays approximately on the following two parallel $[x(t), y(t)]$ trajectories:
\begin{center}
  $m_{2}\gamma x -  m_{1} y + (-1)^{k-1} 2 \textrm{Re}(f_{1}) \pm \left(  m_{1}^{2}+ m_{2}^{2} \right) t=0$.
\end{center}

Reverse-time nonlocal DSII  equation($\gamma^2=-1$;  Here, we take $\gamma=\textrm{i}$ in the following discussion).
\vspace{0.1cm}

(i).  For the first case, if  $\sigma=1$, $r_1=1$, i.e., $|\lambda_{1}|=1$. We obtain the fundamental line rogue wave.    In this case,  $q_{1}(x,y,t)$
approaches a constant background  as $t\rightarrow \pm \infty$. Defining $m_{1}=r_{1}-\sigma r_{1}^{-1}, m_{2}=r_{1}+\sigma r_{1}^{-1}$, then $F_{1}(t)$  and $F(x,y,t)$ becomes:
\begin{eqnarray*}
&& \emph{F}_{1}(t)=  -8 \textrm{i} t  \cos2\varphi_{1} + 2, \\
&& \emph{F}(x,y,t)=\left[ F_{2}(x,y)\right]^2+ 16 t^2 \cos^2 2\varphi_{1} + 1,\\
&& F_{2}(x,y)= 2 x \sin \varphi_{1}  + 2 y \cos \varphi_{1}  - 2 \textrm{i} f_{1}+1.
\end{eqnarray*}
This line rogue wave solution,  if Re$(f_{1})\neq 0$  and (Re$(f_{1}))^2\geq \frac{1}{4}$, then  function $F(x,y,t)$ becomes zero at symmetrical critical time $t_{c}=\pm \frac{\sqrt{4(\textrm{Re}(f_{1}))^2-1}}{4 \cos 2\varphi_{1}}$. This singularity  occurs on the $(x,y)$ plane:
\begin{equation*}
2 x \sin \varphi_{1}  + 2 y \cos \varphi_{1}  + 2 \textrm{Im}(f_{1})+1=0.
\end{equation*}
For other choice in the parameter $f_{1}$, this solution is the nonsingular   line rogue wave.  The line oriented in the $\left( \cos \varphi_{1},  -\sin \varphi_{1} \right)$ direction of the spatial plane, the orientation angle is $-\varphi_{1}$. The width in line rogue wave is angel-independent. For any given time, this solution is a constant along the line with fixed $x \sin \varphi_{1} + y \cos \varphi_{1}$, and approached the constant background 1 with $x \sin \varphi_{1} + y \cos \varphi_{1} \rightarrow \pm\infty$. When  $t\rightarrow \pm\infty$,
the solution $q$ uniformly approaches the constant background $1$ everywhere in the spatial plane.  In the intermediate times, $|q|$ reaches
maximum amplitude  $\frac{4 (\textrm{Re}f_{1})^2+3}{4 (\textrm{Re}f_{1})^2-1}$ (more than three times the background amplitude) at the center  of the line wave at time $t=0$,  which is parameter-dependent.  This line rogue wave is displayed in Fig.7 with parameters chosen as $\varphi_{1}=\frac{\pi }{6}, r_{1}=1, f_1=\frac{1}{4}$.
\begin{figure}[htbp]
   \begin{center}
    \vspace{2.0cm}
   \includegraphics[scale=0.200, bb=0 0 385 567]{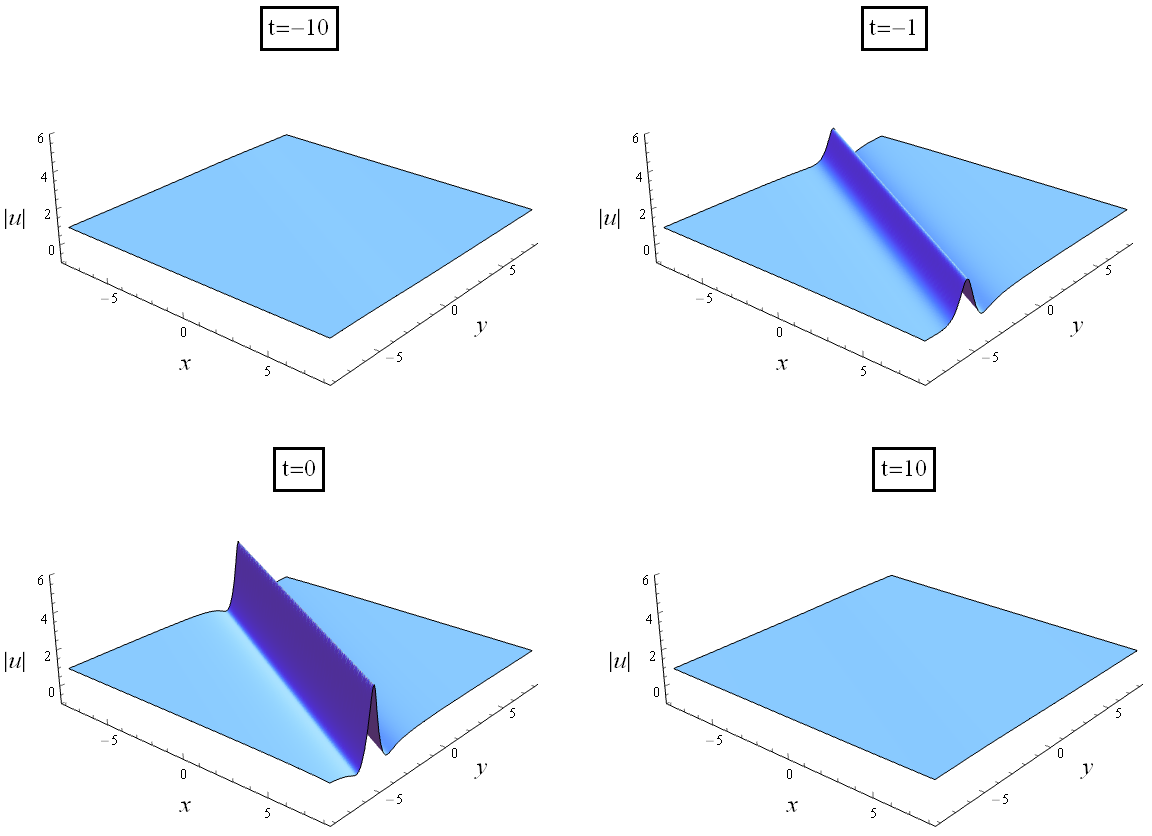} \hspace{5.5cm}
   \caption{ Fundamental line rogue waves in reverse-time nonlocal DS-II equation  with parameters:
   $\varphi_{1}=\frac{\pi }{6}, r_{1}=1, f_1=\frac{1}{4}$.}
   \vspace{-0.5cm}
   \label{Tu7}
   \end{center}
\end{figure}

It is noted that when $\cos 2\varphi_{1}=0$, this line wave is oriented in $45^{\circ}$ or $-45^{\circ}$. In this case, this rational solution is not a rogue wave. Instead, it becomes a stationary line soliton sitting on the constant background.  Moreover, when $\textrm{Re}(f_{1})=0$, this solution also satisfies (\ref{Con-RTsymmetry}) so that $q_{1}(x,y,t)$ admits the local DSII equation\cite{OhtaJKY2013}.

(ii). For the second case, if  $\sigma=-1$ and  $r_1=1$,  it generates the rational travelling waves with:
\begin{eqnarray*}
&& \emph{F}_{1}(t)=  -8 \textrm{i} t  \cos2\varphi_{1} + 2, \\
&& \emph{F}(x,y,t)=-\left[H_{1}(x,y)\right]^2+ 16 t^2 \cos^2 2\varphi_{1} + 1,\\
&& H_{1}(x,y)= 2 x \cos \varphi_{1}  - 2 y \sin \varphi_{1} + 2  f_{1}+ \textrm{i},
\end{eqnarray*}
and the ¡°ridge¡± of the solution lays approximately on the following two parallel $[x(t), y(t)]$ trajectories:
\begin{center}
  $2 x \cos \varphi_{1}  - 2 y \sin \varphi_{1} + 2 \textrm{Re}(f_{1}) \pm 4 t \cos 2\varphi_{1}=0$,
\end{center}
Noticing that the partially $\cal{PT}$-symmetric nonlocal DS system admits the cross-shaped travelling wave solution\cite{JBY2017}.

\subsubsection{ \label{sec:level3} Muliti  rogue waves}
One of the important subclass of non-fundamental rogue waves is the multi-rogue waves, which  describe the interaction between $n$ individual fundamental rogue waves.
These rogue waves are obtained when we take $n>1$ in rational solution (\ref{un+1})-(\ref{wn+1})  with $n$ real parameters $r_{1},...,r_{n}$ or $\varphi_{1}, ..., \varphi_{n}$.  When $t\rightarrow \pm \infty$, the solution $(q, \phi)$ uniformly approaches the constant background $1$  in the entire spatial plane. In the intermediate times, $n$ line rogue waves arise from the constant background, intersect and interact with each other, and then disappear into the background again. For the nonlocal DSI equation, its multi rogue-wave solution consists of  $n$ separate line rogue waves in the far field of the spatial plane. However, in the near field, the wavefronts of the rogue wave solution are no longer lines, and there would be some interesting curvy wave patterns.
For the nonlocal DSII equation, the multi rogue-waves appear in the form of exploding rogue waves, which develop singularities under suitable choice of parameters.

To demonstrate these multi rogue-wave solutions in the reverse-time nonlocal DS system, we first consider the case of $n=2$. In this case, two rogue-wave solutions for this reverse-time nonlocal DS system are uniformly contained  in only one expression, where $\lambda_{1}$, $\lambda_{2}$, $\varphi_{1}$, $\varphi_{2}$ are free real parameters, $f_{1}$, $f_{2}$ are free complex parameters.
\begin{figure}[htb]
   \begin{center}
   \vspace{3.0cm}
   \includegraphics[scale=0.200, bb=0 0 385 567]{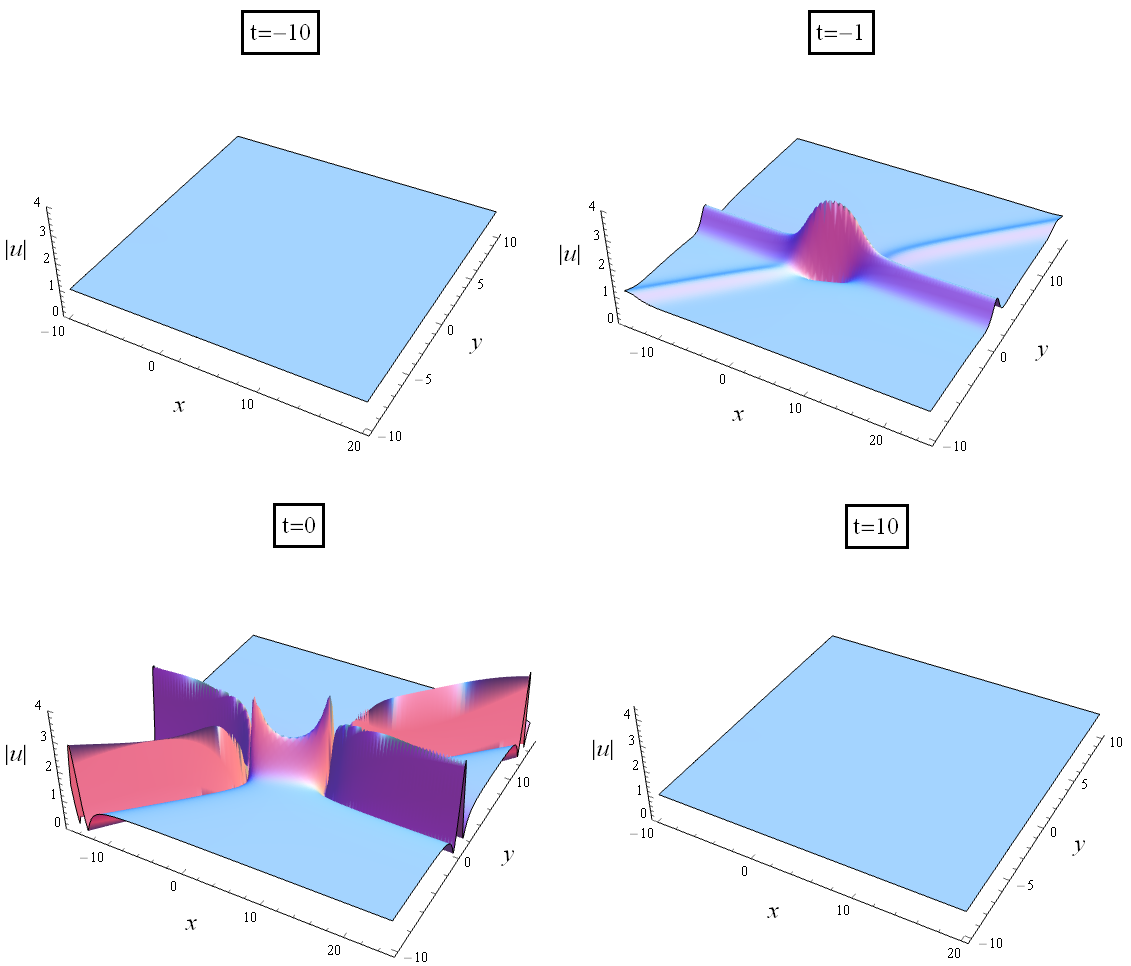} \hspace{5.5cm}
   \caption{ A two-rogue waves in reverse-time nonlocal DS-I equation  with parameters (\ref{twoRWSpara1}). }
   \vspace{-0.25cm}
   \label{Tu8}
   \end{center}
\end{figure}

When $\gamma=1$, i.e., for the nonlocal DSI equation, a two-rogue-wave solution with parameters
\begin{eqnarray}\label{twoRWSpara1}
\lambda_1=1,\ \ \lambda_2=2,\ \ {f}_1=0.05,\ \ f_2=-0.01
\end{eqnarray}
is shown in Fig. 8.

Firstly, these two line rogue waves,  arising from the constant background,  possess higher amplitude  in the intersection region at $t=-1$.  Afterwards, these higher amplitudes in the intersection region fade, then in the far field, two line rogue solutions rise to higher amplitude at $t=0$.  Afterwards, the solution goes back to the constant background again at large times (see the $t=10$ panel). During this process, the interaction between two fundamental line rogue waves does not generate very high peaks. Actually, the maximum value of solution $|q|$ does not exceed $4$ for all times (i.e., four times the constant background). It is noted that this kind of wave pattern is first found in the local DSI equation\cite{OhtaJKY2012}.  Interestingly, if $f_{1}$, $ f_{2}$ in (\ref{twoRWSpara1}) are  purely imaginary numbers, we have $q^*(x,y,t)=q(x,y,-t)$.  In that case, this  two-rogue-wave solution also admits the local DSI equation.

However, if we choose the value of real parameter $r_{1}$ not to be one. For example, if we set
\begin{eqnarray}\label{twoRWSpara2}
\lambda_1=1/2,\ \ \lambda_2=2,\ \ {f}_1=0.05,\ \ f_2=-0.01,
\end{eqnarray}
as what is shown in Fig.9,
\begin{figure}[htb]
   \begin{center}
   \vspace{3.0cm}
   \includegraphics[scale=0.200, bb=0 0 385 567]{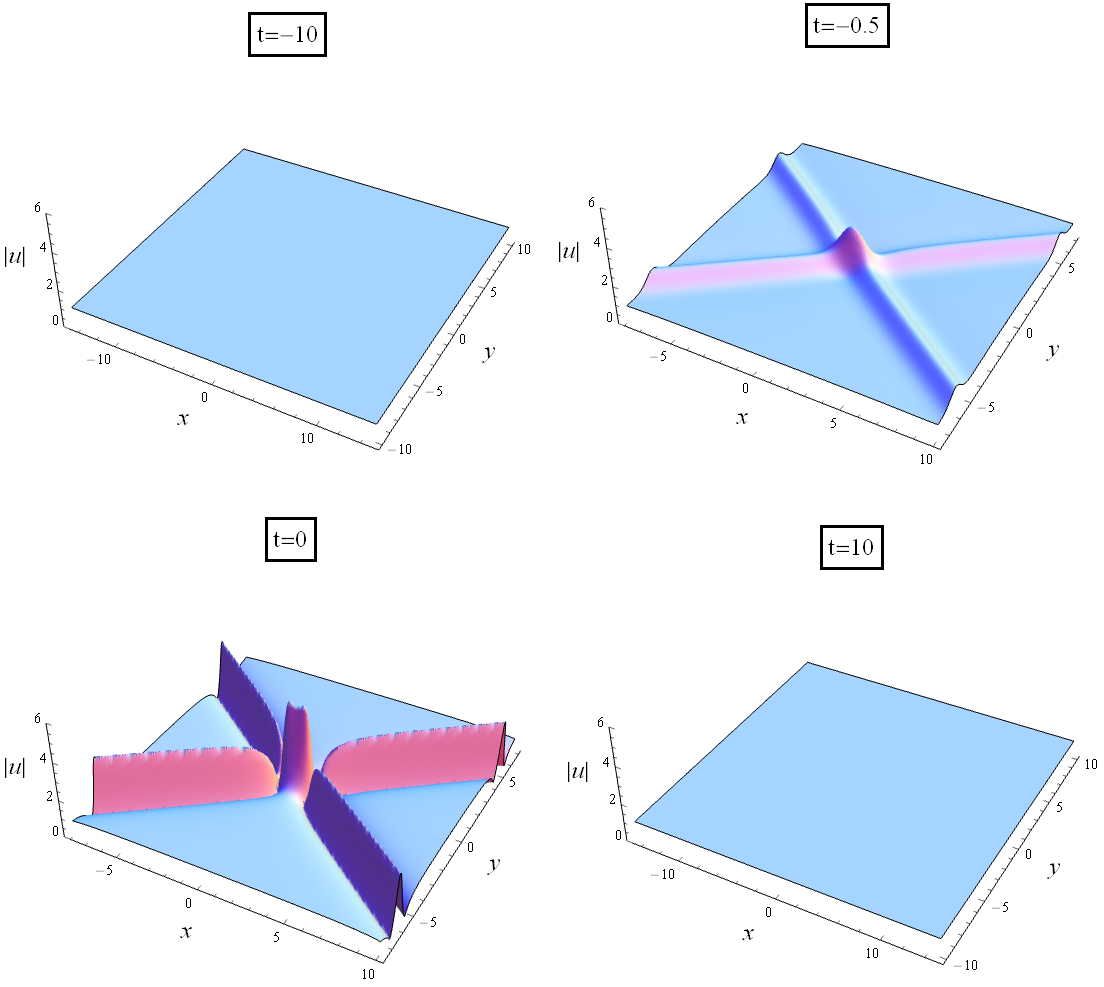} \hspace{5.5cm}
   \caption{ A two-rogue waves which generate higher amplitude in reverse-time nonlocal DS-I equation  with parameters (\ref{twoRWSpara2}). }
   \vspace{-0.5cm}
   \label{Tu3}
   \end{center}
\end{figure}
the maximum value of solution $|q|$  becomes higher and exceeds the value $4$,  which is different from the previous pattern.

Thus, for larger $n$, multi-rogue waves in the reverse-time nonlocal DSI equation have qualitatively similar behaviors, more fundamental line rogue waves will arise
and interact with each other, and more complicated wave fronts will emerge in the interaction region. For example, with $n=3$
and parameter choices the corresponding solution is shown in Fig.10.

As can be seen, the transient solution patterns become more complicated.
Also, as we could see, wether the maximum value of solution $|q|$ stays below or beyond $4$  for all times depends on the choice of the real parameters $\lambda_{i}$. For example, the parameters for Fig.10 are chosen as:
\begin{eqnarray}\label{threeRWSpara1}
\lambda_1=1,\ \lambda_2=1.5,\ \lambda_3=2,\  f_1=0.05,\ f_2=0,\ f_{3}=-0.05. \nonumber \\
\
\end{eqnarray}
In this case, the transient solution patterns become more complicated. But the maximum value of this solution $|q|$ still stays below 4 in all times. However, if one takes $\lambda_{1}=1/3$ in above parameters,  this interaction would be able to create very high amplitude.
\begin{figure}[htbp]
   \begin{center}
    \vspace{2.0cm}
   \includegraphics[scale=0.213, bb=0 0 385 567]{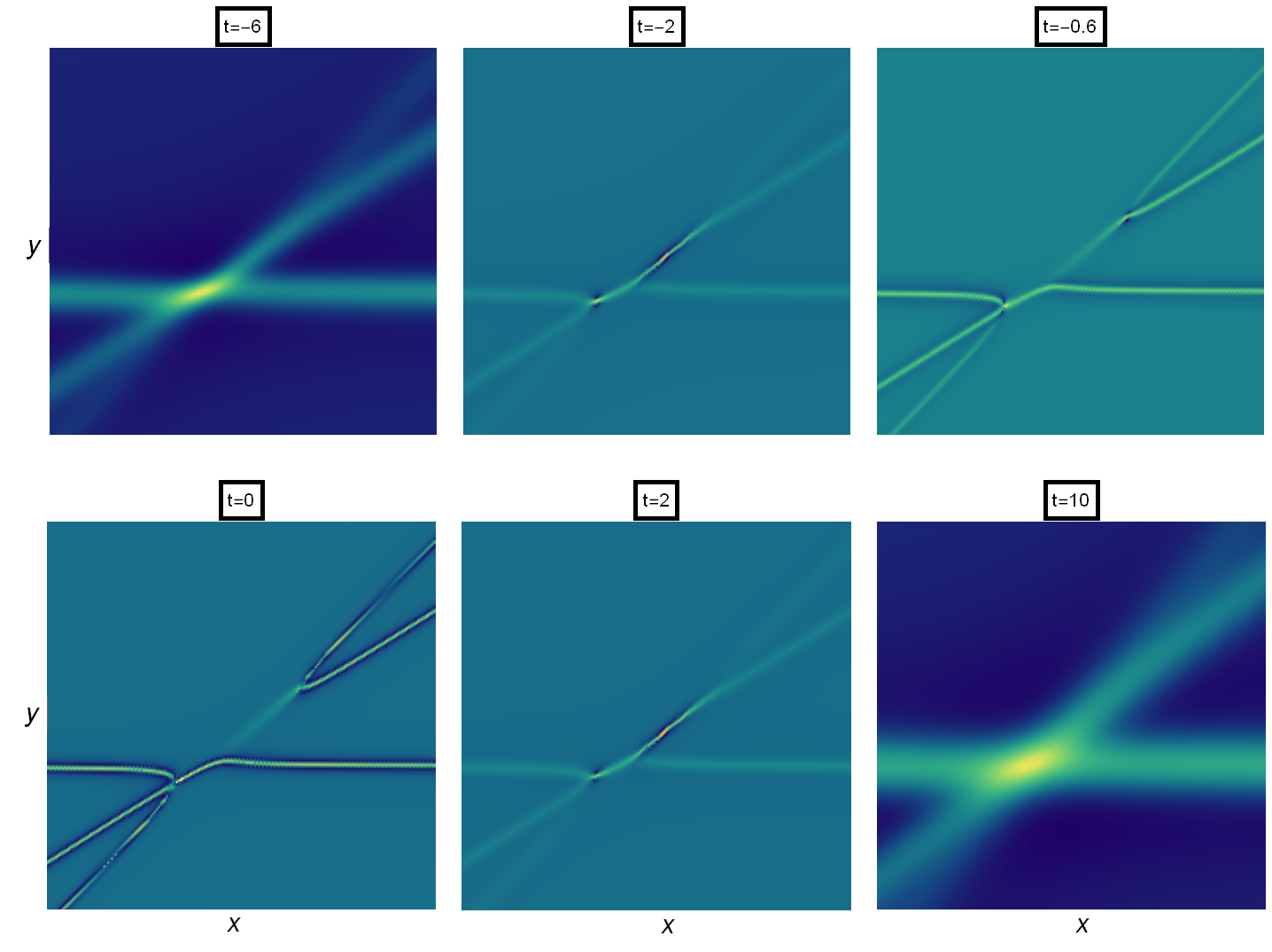} \hspace{5.25cm}
   \caption{A three-rogue waves in reverse-time nonlocal DS-I equation  with parameters (\ref{threeRWSpara1}). }
   \vspace{-0.25cm}
   \label{Tu10}
   \end{center}
\end{figure}

Next, when $\gamma=1$, we find some exploding rogue-wave solutions  for the reverse-time nonlocal DSII equation. These exploding rogue waves, which arise from the constant background 1, can blow up to infinity in a finite time interval at isolated spatial locations under certain parameter conditions. In this case, the solutions go to a constant background, $q \rightarrow1,\ \phi \rightarrow\sigma$, as $t \rightarrow -\infty$. To demonstrate, we consider a two-rogue-wave solution whose expression are given by taking $n=2$ in (\ref{un+1})-(\ref{wn+1}). Choosing parameters as
\begin{eqnarray}\label{ExplodingRWpara}
\lambda_1=1,\ \ \lambda_2= \textrm{i},\ \ {f}_1=0.05,\ \ f_2=0.025.
\end{eqnarray}
This exploding rogue-wave solution is displayed in Fig.11.
\begin{figure}[htb]
   \begin{center}
   \vspace{2.5cm}
   \includegraphics[scale=0.200, bb=0 0 385 567]{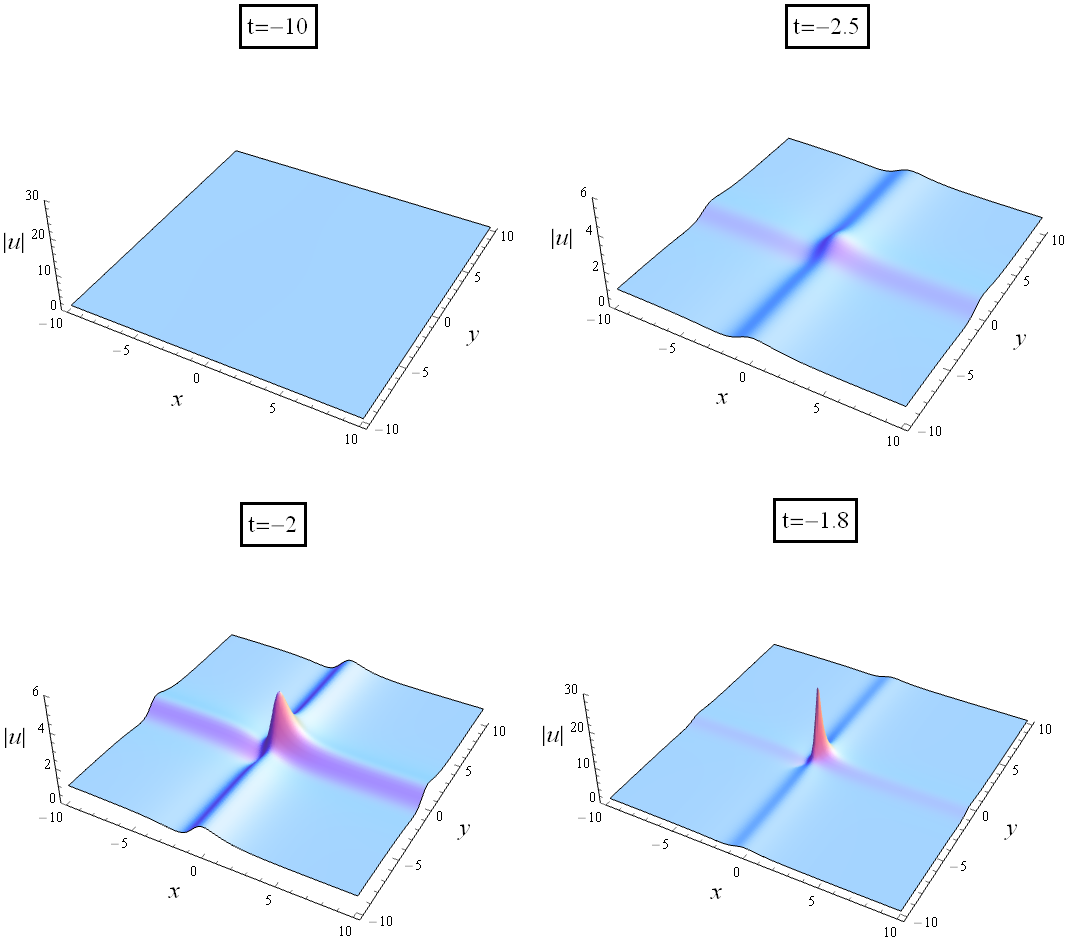} \hspace{5.5cm}
   \caption{The exploding rogue waves in reverse-time nonlocal DS-II equation  with parameters (\ref{ExplodingRWpara}).}
   \vspace{-0.25cm}
   \label{Tu11}
   \end{center}
\end{figure}

One can see a cross-shape wave firstly appears in the intermediate times (see the $t=-2.5$, $t=-2$ panels), which describes the nonlinear interaction between two fundamental line rogue waves: one of the line  oriented along the  $x$-direction (corresponding to parameter $\lambda_{1}$), the other line oriented along the $y$-direction (corresponding to parameter $\lambda_{2}$). Afterwards, the peak amplitude in this cross-shape rogue wave becomes much higher instantaneously.  In Fig.11, we plotted solution up to time $t=-1.8$, shortly before the blowing-up.  One can see the maximum amplitude for this rogue wave solution  becomes extremely high.

For the local DSII equation, multi and higher-order collapsing wave solutions with constant boundary conditions have been derived in \cite{OhtaJKY2013}. Moreover, for the partially $\mathcal{PT}$-symmetric nonlocal system, similar exploding rogue waves also exist, where the blowups can occur on an entire hyperbola of the spatial plane.

\subsubsection{ \label{sec:level3} High-order rogue waves}

Another important subclass of non-fundamental rogue waves is the higher-order rogue waves, which are obtained from the higher-order rational solution  with a real
value of $\lambda_{1}$.  For example, if $n=2$ in (\ref{gDTDS-IIQN})-(\ref{gDTDS-IIwN}), and we get the second-order rogue-wave solution. To demonstrate these dynamic behaviours, we consider above 2-nd order rogue waves with parameters
\begin{eqnarray}\label{HigherRWpara}
\gamma=1,\ \sigma=1,\ \lambda_1=1,
\end{eqnarray}
while $f_{1}$ is a free complex parameter. Thus we get
\begin{eqnarray}\label{High-orderRws}
q_{1}(x,y,t)=\frac{g_{1}}{h_{1}},
\end{eqnarray}
where $\zeta(y):=y^2+y+\frac{1}{2}$,
\begin{eqnarray*}
&&h_{1}=\left(8 t^2-2 x-2\zeta \right)^2+ \zeta \left(64 t^2+4\right)+4 \textrm{i} f_1 \left(2 y+1-\textrm{i} f_1\right) \left[-(2 y+1-\textrm{i} f_1 )^2 - 8t^2 + 2\zeta -2x - 2\right],\\
&&g_{1}=h_{1}-2\left( 1+ 4 \textrm{i} t\right) \left[\left(-2\textrm{i} f_1+2y+1\right)^2-\left(1+ 4 \textrm{i} t\right)^2-4 x\right].
\end{eqnarray*}
For solution (\ref{High-orderRws}), if $f_{1}$ is taken as a purely imaginary number, then denominator function $h_{1}$ is real, thus  relation (\ref{Con-RTsymmetry}) is  satisfied. However, if $f_{1}$ is chosen as real number, then $h_{1}$ is complex, in that case,  this solution do not satisfy  the local DSI equation.  By choosing suitable value in $f_{1}$, this solution can be   nonsingular.

An interesting dynamic behaviour  for this  solution is that,  these higher-order rogue waves do not uniformly approach the constant background as $t\rightarrow \pm \infty$. Instead, only parts of their wave structures approaches background as $t\rightarrow \pm \infty$.  For instance,  if we  we set  $f_{1}=0.05$.  When $|t|\gg1$, this solution becomes a  localized lump sitting on the constant background 1 (see the $t=\pm6$ panels). And this lump disappears as $t\rightarrow 0$. At the same time, a parabola-shaped rogue wave generates from the background. Moreover, when $t=0$, this parabola is approximatively located at
\begin{center}
$x=-y^2-y-\frac{1}{2}$.
\end{center}

Visually, the solution displayed in Fig.12  can be described as an incoming lump being reflected back by the appearance of a parabola-shaped rogue wave. This interesting pattern is firstly obtained via the bilinear method  in ref~\cite{OhtaJKY2012}. It is indeed surprising that a similar pattern can be produced in this reverse-time nonlocal DSI equation, although the expression for this solution can be different.
\begin{figure}[htb]
   \begin{center}
   \vspace{3.0cm}
   \includegraphics[scale=0.160, bb=0 0 385 567]{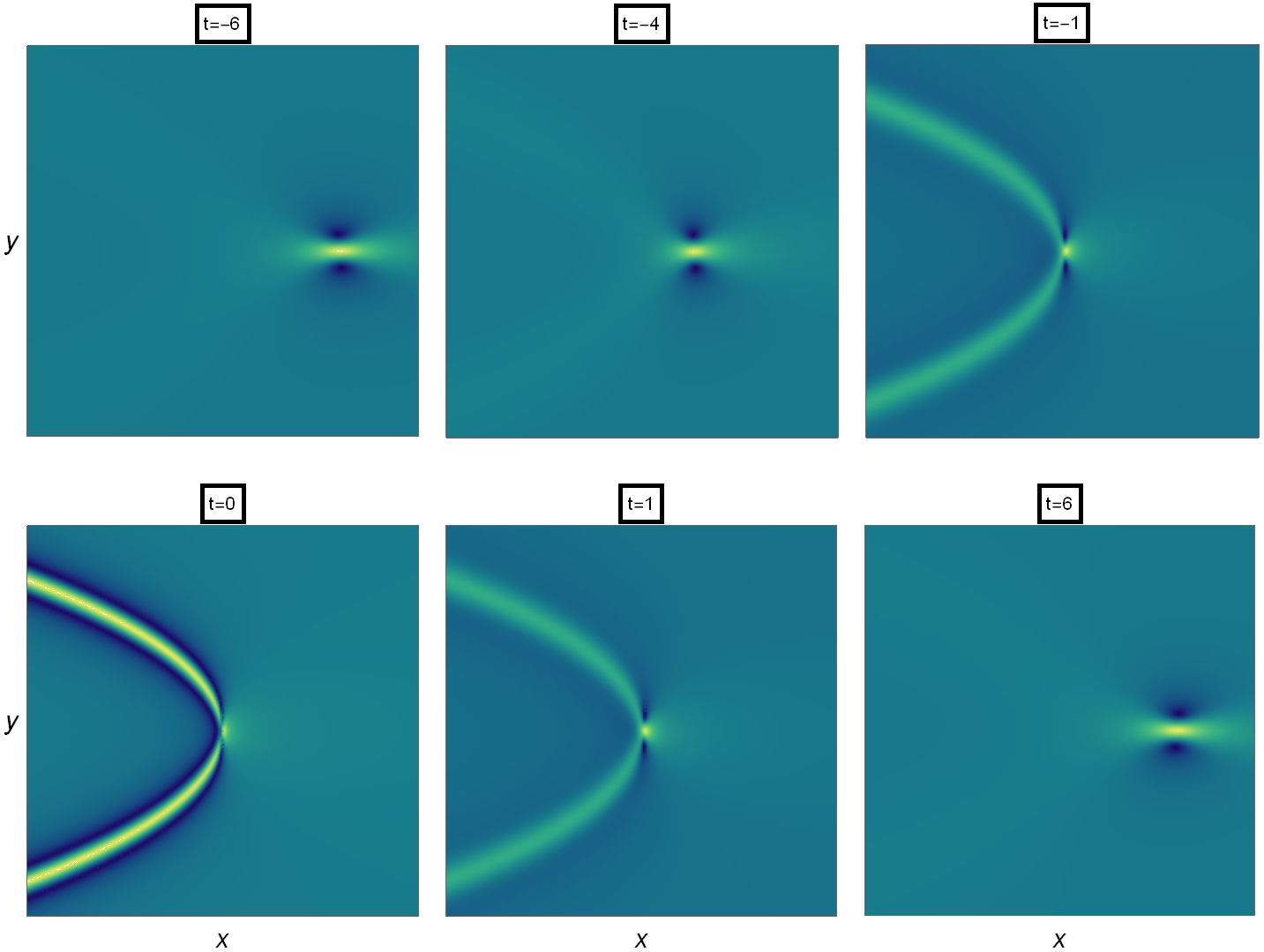} \hspace{5.5cm}
   \caption{ The second order rogue waves in reverse-time nonlocal DS-I equation  with parameters (\ref{HigherRWpara}). }
   \vspace{-0.5cm}
   \label{Tu4}
   \end{center}
\end{figure}

However, when $\gamma^2=-1$, we only derive some higher-order rational solutions with almost full-time singularities.   Hence,  it is still unknown whether  there exists nonsingular or finite-time blowing-up  high-order rogue waves in the reverse time nonlocal DS-II equation.

\section{Summary and discussion}
In this article,    general rogue waves have been derived for the reverse time nonlocal NLS equation (\ref{e:RTNLS}) and the reverse time nonlocal DSI and DSII equations (\ref{PTDS})-(\ref{PTDSS}) by  Darboux transformation method under certain symmetry reductions.  It is  interesting  to show that  a new unified binary DT has been constructed for this reverse time nonlocal DS system. Thus, the rogue-wave solution in reverse time nonlocal DSI and DSII equation  can be written  in a unified formula.

New  dynamics of rogue waves is further explored.  It is shown the  (1+1)-dimensional fundamental rogue waves  can be bounded for both $x$  and $t$, or develop collapsing singularities.  It is also shown that the (1+2)-dimensional fundamental line rogue waves can be bounded for all space and time, or have finite-time blowing-ups. All these types depend on the  values of  free parameters contained in the solution. In addition, the dynamics patterns in the multi- and higher-order rogue waves exhibits  richer structures, most of which are found  in the nonlocal integrable equations for the first time. For example,  the (1+1)-dimensional higher-order rogue waves exhibit more hybrid of  collapsing and non-collapsing peaks, arranging in triangular, pentagon, circular and other erotic patterns.  The (1+2)-dimensional multi-rogue waves  describe the interactions between several fundamental line rogue waves,  and  some curvy wave patterns with higher amplitudes  appear due to the interaction. However, for the (1+2)-dimensional higher-order rogue waves,  only part of the wave structure rises from the constant background and then retreats back to it, which  possesses the parabolas-like shapes.  While the other part of the wave structure comes from far distance as a localized lump, which interacts with the rogue waves in the near field  and then reflects  back  to the large distance again. These rogue-wave  to the nonlocal equations generalize the rogue waves of the local equation into the reverse-time satiation, which could play a role in the physical understanding of rogue water waves in the ocean.  In addition, the DT reduction method in our paper can be generalized to look for  rogue waves in other types of integrable nonlocal equations, for example, the reverse-space,  or the reverse-space-time  nonlocal equations\cite{AblowitzMussSAPM}.

\section*{Acknowledgment}
This project is supported by the  National Natural Science Foundation of China
(No.11675054 and 11435005), Global Change Research Program of China (No.2015CB953904), and Shanghai Collaborative Innovation Center of Trustworthy Software for Internet of Things (No. ZF1213).

\end{document}